\documentclass[varenna,seceqno]{cimento}
\input BoxedEPS
\SetRokickiEPSFSpecial  
\HideDisplacementBoxes

\def\rr{\right\rangle}
\def\ll{\left\langle}
\def\underarrow#1{\mathrel{\mathop{\longrightarrow}\limits_{#1}}}
\def\svec#1{\skew{-2}\vec#1}
\def\ccslash#1{\not{\!\!#1}}
\def\cslash#1{\not{\!#1}}
\def\ssquare{\kern1pt\vbox{\hrule height .6pt\hbox{\vrule width .6pt\hskip 3pt
   \vbox{\vskip 6pt}\hskip 3pt\vrule width 0.6pt}\hrule height 0.6pt}\kern1pt}

\def\sq{\mbox{\scriptsize$\sqcap$\llap{$\sqcup$}}}
\def\Sq{\mbox{$\sqcap$\llap{$\sqcup$}}}

\def\fracs#1#2{{\textstyle\frac#1#2}}
\def\xx#1#2{x_{#1},\ldots,x_{#2}}
\def\Dx#1#2{{\cal D}(\xx{#1}{#2})}

\def\Tr{\mathop{\rm Tr}}
\def\tr{\mathop{\rm tr}}
\def\Re{\mathop{\rm Re}}
\def\notp{/\llap{$p$}}
\def\notA{\rlap{/}A}

\def\be{\begin{equation}}
\def\ee{\end{equation}}
\def\bea{\begin{eqnarray}}
\def\eea{\end{eqnarray}}
\makeatletter
\def\@copyright@{{\null}}
\makeatother

\title{Understanding Hadron Structure Using Lattice QCD}
\author{J.W. Negele}
\institute{Center for Theoretical Physics\\
Laboratory for Nuclear Science
and Department of Physics\\ 
Massachusetts Institute of Technology, Cambridge
MA 02139}

\begin{document}

\maketitle

\vspace*{-1.33in}

{\small \paragraph{Abstract}
  An   elementary introduction is presented to the study of hadron structure
  using lattice QCD\null. Following a brief review of relevant aspects of path integrals, the
  discrete lattice path integral is presented for gluon and quark fields and used to
  calculate physical observables. Essential aspects of instanton physics are reviewed, and
  it is shown how the instanton content is extracted from lattice gluon configurations.
  Finally, both comparison of results including all gluons with those including only
  instantons and the study of quark zero modes associated with instantons and their
  contributions to hadronic observables are used to show the dominant role of gluons in
  hadron structure. \hfill \footnotesize MIT CTP\# 2701\quad hep-lat/9804017\par}

\section{Introduction and Motivation}\label{JWN:sec:1}
Although a quarter of a century  has passed since the experimental discovery of
quarks and the formulation of QCD, we are only now beginning to understand the
essential physics of the structure of light hadrons.  To truly understand hadron
structure, one must solve rather  than model QCD, and the only known means to do
so is the numerical solution of lattice field theory.  But obtaining accurate numerical
results for observables from a computer is not enough --- we also need to obtain
physical insight.  Hence, our strategy is to use numerical evaluation of the QCD path
integral on a lattice to identify the configurations that dominate the action as well
as to calculate observables.  In recent years, the algorithms and techniques of
lattice QCD and the performance of massively  parallel computers have developed to
the point that we are now on the threshold of reliable, quantitative calculations of
QCD observables.  Furthermore, there is strong evidence from lattice calculations
that the topological excitations of the gluon field corresponding in the semiclassical
limit to instantons play a dominant role in the structure of light hadrons.  The
purpose of these lectures is to describe at an elementary level the basic elements of
lattice QCD and how numerical solution of QCD on a lattice is elucidating the role of
instantons in light hadrons.

To appreciate the significance of the current lattice results, it is useful to recall the wide
range of disparate physical pictures that have arisen from the different QCD inspired
models introduced to model hadrons.   For example, non-relativistic quark models focus
on constituent quarks interacting via an adiabatic potential.  Bag models postulate a
region in which relativistic current quarks are confined and interact by gluon
exchange.  Motivated by large $N_c$ arguments, Skyrme models describe the nucleon
as a topological soliton built out of $q \bar{q}$ pairs. Finally instanton models
emphasize the role of topological excitations of the vacuum and of the quark zero
modes associated with these excitations.  

Unfortunately, phenomenology has proven inconclusive in determining which, if any, of
these fundamentally different pictures describes the essential physics of light hadrons,
since each of the models is rich enough that with sufficient embellishment it can be
made to fit much of the data.  Whereas perturbative QCD has proven extremely useful
in extracting quark and gluon structure functions from high energy scattering
experiments, it is inadequate to understand their origin.  Hence, it is necessary to turn
to nonperturbative solution of QCD on the lattice.

The physical picture that arises from this work corresponds closely to the physical
arguments and instanton models of Shuryak and others~\cite{R:JN:03,R:JN:04,R:JN:05} in
which the zero modes associated with instantons produce localized quark states, and
quark propagation proceeds primarily by hopping between these states.    Thus, QCD
with light quarks is unique among the many-body systems with which we are familiar
in the sense that the quanta generating the interactions cannot be subsumed into a
potential but rather participate as essential dynamical degrees of freedom. In atoms, for
example, photons play a negligible dynamical role, and to an excellent approximation
may be subsumed into the static Coulomb potential.  In nuclei, mesons play a minor
dynamical role, and to a good approximation nuclear structure maybe described in
terms of two- and three-body nuclear forces. Indeed, experimentalists need to work
very hard and pick their cases carefully to observe any effects of meson exchange
currents. And in heavy quark systems, much of the physics of $c\bar{c}$ and
$b\bar{b}$ bound states may be understood by subsuming the gluons into an adiabatic
potential with Coulombic and confining behavior. The fact that nucleons are completely
different in that gluons are crucial dynamical degrees of freedom is not
entirely unexpected.  Indeed,  from perturbative QCD, we already know by the work of
Gross and Wilczek~\cite{R:JN:01} and Hoodbhoy, Ji, and Tang~\cite{R:JN:02} that
approximately half (${16}/{3n_f}$ to be precise, where $n_f$ is the number of active
flavors and equals 5 below the top quark mass) of the momentum and angular
momentum comes from glue in the limit of high $Q^2$.  Furthermore, experiment tells
us that this behavior continues down to non-perturbative scales of the order of several
GeV$^2$.

The discussion of lattice QCD will be based on four key, underlying ideas.   The first is the use of
path integrals.  One of the great contributions of Feynman to theoretical
physics was the formulation of quantum mechanics in terms of path integrals,
which provides both a physical picture of quantum evolution in terms of sums
of time histories and a powerful computational framework.  For the present
application, we will make use of the fact that the path integral eliminates
the non-commuting operators of quantum mechanics or field theory by
introducing an integral over an additional continuous variable, and thus
effectively reduces the problem of quadrature.
 
The second major idea is the introduction of Euclidean time.  The basic idea
is to write $|\psi\rangle =e^{-\beta H} |\phi\rangle$, where $e^{-\beta H}$
acts as a filter to project the ground state $|\psi\rangle$ out of an
arbitrary state $|\phi\rangle$ having the desired set of quantum numbers, so
the continuous variable in the path integral is imaginary or Euclidean time.
The resulting theory has important connections with statistical mechanics.  In
the case in which one sums over a complete set of states and calculates the
trace $\Tr e^{-\beta H}$, one is solving field theory at finite temperature
and $\beta$ corresponds to the physical inverse temperature.  The
corresponding path integral has the structure of classical statistical
mechanics in $d+1$ dimensions.  Many familiar ideas from statistical physics
concerning critical behavior, order parameters, and Landau's theory of phase
transition turn out to be useful.
 
The third principal idea is lattice regularization, which replaces continuum
field theory by a finite quantum many-body problem on a lattice.  For any
finite lattice spacing $a$, the maximum momentum which can arise on the
lattice is $p_{\rm max}\sim {\pi\over a}$, so that the lattice effectively
imposes a momentum cutoff of order $p_{\rm max}$ which goes to infinity as the
lattice spacing goes to zero.  One of the great accomplishments in recent years has been
the use of renormalization group arguments and other techniques to provide convincing
approximation to\label{P:M:JN:1}
the underlying continuum theory.\cite{M:JN:1}
 
The last key idea is the use of stochastic, or Monte Carlo, methods to
evaluate the lattice path integrals.  One should note at the outset that the
common misnomer of Monte Carlo ``simulations'' is quite misleading.  In fact,
we are not simulating anything.  Rather, we are solving an equation in the
same sense as one always uses numerical analysis to solve equations.  That is,
one first selects a desired level of precision, and then using appropriate
theorems, determines an algorithm and a number of independent samples which
yields that precision.  

The outline of these lectures is as follows.  Following this introduction, aspects of
path integrals relevant to lattice QCD are reviewed in Section~\ref{JWN:sec:2}.  The
basic ideas of lattice QCD are presented  beginning with the pure gluon sector in
Section~\ref{JWN:sec:3} and then adding quarks  in
Section~\ref{JWN:sec:4}.  The role on instantons in light hadrons is discussed in
the final section.  Following an overview of instantons, it is shown how the
instanton content is extracted from lattice gluon configurations.  Results including
all gluon contributions are compared with those including only instantons to
provide one indication of the dominant role of instantons.  Finally, direct calculation
of quark zero modes, observation of quark localization for these modes at the
locations of instantons, and demonstration that these modes dominate the rho and
pion contributions to vector and pseudoscalar correlation functions provide
additional indications of the role of instantons in hadron structure.

For readers who wish to go beyond the scope of the present lecture, I
recommend several basic references.  Much of the background material is
discussed in more detail in a text
co-authored with Orland.\cite{R:JN:10} In
particular, the reader is referred to Chapter 1 for treatment of coherent
states and Grassmann variables, Chapter 2 for discussion of path integrals, and
Chapter 8 for a detailed explanation of stochastic methods.  A terse
introduction to lattice gauge theory is provided by Creutz~\cite{R:JN:06} and more
details may be found in the reprint volume edited by Rebbi~\cite{M:JN:xx} which
includes all the key articles through 1983 and in comprehensive texts written by
Rothe~\cite{M:JN:2} and by Montvay and M\"unster~\cite{M:JN:3}.  Up-to-date reviews
of recent results may be found in the proceedings of the yearly lattice conferences
published in {\it Nuclear Physics B\/} Proceedings Supplements.

\setcounter{equation}{0}
\section{Path Integrals}\label{JWN:sec:2}

\subsection{Feynman Path Integral}

The basic idea of the path integral is illustrated by considering the Feynman
path integral for a single degree of freedom.  The evolution operator
$e^{-iHt}$ is broken up into a large number of ``time slices'' separated by
time interval $\epsilon$, and a complete set of states is inserted between
each interval
\begin{equation}
e^{-iHt} = e^{-iH\epsilon}\int dx_n|x_n\rangle\langle x_n|\,e^{-iH\epsilon}
\int dx_{k-1}|x_{n-1}\rangle\langle x_{n-1}| e^{-iH\epsilon}\cdots
\label{eq:2.1}
\end{equation}
Then, the non-commutativity of the kinetic and potential energy operators is
treated by the following approximation which becomes exact in the limit
$\epsilon\to 0$
\begin{eqnarray}
&& \hskip-1.5cm \langle x_{k+1}|  e^{-i\epsilon\left({\hat p^2\over 2m} +
V(\hat x)\right)} | x_k \rangle \sim \langle x_{k+1}|
e^{-i\epsilon{\hat p^2\over 2m}} \int dp|p\rangle \langle p| e^{- i
\epsilon V(\hat x)}| x_k\rangle \nonumber \\
&=& \int dp\,e^{ip (x_{k+1} - x_k) - i\epsilon{p^2\over 2m} - i\epsilon V(x_k)}
\label{eq:2.2} \\
&=& \sqrt{{2m\pi\over \epsilon}} \,e^{i\epsilon\sum_k \left[ {m\over
2}
\left( {x_{k+1} - x_k\over\epsilon}\right)^2 - V(x_k)\right] } +{\cal O}(\epsilon^2)
\nonumber
\end{eqnarray}
Hence the evolution operator may be expressed as the sum over all
paths of the exponential of the classical action
\begin{eqnarray}
\ll x_f\left| e^{-iHt}\right| x_i\rr
 &=& \int \Dx1n \,  e^{i\epsilon\sum_k\left[{m\over 2}\left(
{x_{k+1}-x_k\over\epsilon}\right)^2 - V(x_k)\right]} \nonumber \\
&&\to \int^{x(t) = x_f}_{x(0) = x_i}
 \Dx1n \,e^{i\epsilon S_{\rm classical}\left(
x(t)\right) }
\label{eq:2.3}
\end{eqnarray}
The quantum mechanics of the non-commuting operators $\hat x$ and $\hat p$ has
thus been represented by an ordinary integral over an additional time
variable.  This result may be generalized to many degrees of freedom as
follows
\begin{equation}
\mkern24mu\ll x^f_1\cdots x^f_N\left| e^{-iHt}\right| x^i_1\cdots x^i_N\rr =
\int^{x^f_1\cdots x^f_N}_{x^i_1\cdots x^i_N}
\!\! e^{i\epsilon\sum_k\Bigl[ \sum_i{m\over 2} \bigl(
{x^{k+1}_{i} - x^k_i\over\epsilon}\bigr)^2 - {1\over 2} \sum_{ij} v\left(
x^k_i-x^k_j\right) \Bigr] } 
\label{eq:2.4}
\end{equation}
where a complete set of states
$|x_1\cdots x_N\rangle$ $\langle x_1\cdots x_N|$,
is inserted at each time slice.
 
One important property of the path integral
is that a time-ordered product is represented as follows:
\begin{eqnarray}
T\,{\cal O}(t_1) {\cal O}(t_2)\,e^{-i\int^T_0 dtH(t)} &=&
e^{-iH(T-t_2)}{\cal O} (t_2) \,e^{-iH(t_2-t_1)} {\cal O} (t_1)\,e^{-iH(t_1-0)}
\nonumber \\
&&\to \int \Dx1n \, e^{iS(\xx1n)} {\cal O}(x_{k_2})
{\cal O} (x_{k_1})  
\label{eq:2.5}
\end{eqnarray}
Hence, any path integral composed of $e^{iS}$ and a sequence of operators
automatically corresponds to a time-ordered product.
 
The classical limit is obtained by including the factors of $\hbar$ which have
been suppressed thus far and applying the stationary phase approximation
\begin{equation}
\int {\cal D}(x)\,e^{{i\over\hbar}S(x)} \underarrow{\rm SPA}
e^{{i\over\hbar} S(x_{\rm cl})}
\left({1\over \sqrt{ \det \left( m{d^2 \over dt^2} + V'' \left(
x_0(t)\right)\right)}}
+{\cal O}(\hbar)\right) 
\label{eq:2.6}
\end{equation}
in which case the path integral represents the sum of all quadratic
fluctuations around the classical path.
 
It is important to note that there is nothing sacred about the physical time,
and any continuous variable may be ``sliced'' to treat the non-commutativity of
$\hat x$ and $\hat p$.  A common case is the Euclidean path integral, in which
real time is replaced by imaginary time or temperature, with the result
\begin{equation}
e^{-\beta H} = \prod e^{-\epsilon H}\Longrightarrow \int {\cal
D}(x)\,e^{-\sum_k\epsilon\left[ {m\over 2} \left( {x_{k+1}-x_k\over
\epsilon}\right)^2 + V(x_k)\right]} 
\label{eq:2.7}
\end{equation}
In this analytic continuation in which $it\to\tau$,
the Lagrangian is effectively replaced by the
Hamiltonian in the exponent
\begin{equation}
\int dt\left[ {m\over 2} \dot x^2 - V\right] \underarrow{it\to\tau} \int
d\tau\left[ {m\over 2} \dot x^2 + V\right] 
\label{eq:2.8}
\end{equation}
Salient properties of this Euclidean path integral are the fact that it is
purely real, it has a well-defined measure, the Wiener measure, and it has the
structure of the partition function of statistical mechanics with one extra
dimension.
 
The boundary conditions on the path integral are specified by the specific
matrix element or elements under consideration.  For example, the
thermodynamic trace has the form
\begin{equation}
\Tr e^{-\beta H} = \int dx\ll x\left| e^{-\beta H} \right| x\rr = \int {\cal
D}(x_0, \xx1n) e^{-S(\xx0n)} 
\label{eq:2.9a}
\end{equation}
where
\begin{eqnarray}
S(\xx0n) = \epsilon &\left[ {m\over2} {\left( x_0 - x_n\right)^2\over
\epsilon^2} + V(x_n) + {m\over 2}  {\left( x_n-x_{n-1}\right)^2\over\epsilon} +
V(x_{n-1}) \right. \nonumber \\
&\left.+ \cdots + {m\over 2} {\left( x_1-x_0\right)^2\over\epsilon} +
V(x_0) \right]
\label{eq:2.9b}
\end{eqnarray}
and thus has periodic boundary conditions. For specific matrix elements
however, we obtain the alternative form
\begin{equation}\ll \phi_f\left| e^{-\beta H} \right| \phi_i\rr = \int {\cal D} \left(x_0,
\xx1n, x_{n+1}\right)e^{-S(\xx0{n+1})} 
\label{eq:2.10a}
\end{equation}
where
\begin{eqnarray}
S(\xx0{n+1}) &=& - \ln \phi_f (x_{n+1}) +  \epsilon \biggl[ {m\over 2} \left(
{x_{n+1} - x_n\over \epsilon}\right)^2 + V(x_n) +\cdots \nonumber \\
&&{}+ {m\over 2} \left( {x_1 - x_0\over \epsilon}\right)^2 + V(x_0) \biggr] - \ln \phi_i
(x_0) \ \ . 
\label{eq:2.10b}
\end{eqnarray}

\subsection{Scalar Field Theory}

Using this knowledge of the Feynman path integral, it is now easy to
generalize to scalar field theory on a lattice.  Let the continuum coordinate
${\svec r}$ be replaced by discrete lattice coordinates ${\svec
n}\equiv(n_1,n_2,n_3)$ where the $n_i$ are integers and lengths will be
understood to be in units of the lattice spacing $a$.  Then one simply views
the lattice field theory as a quantum many-body problem where the canonical
coordinate and momentum operators $\hat x_i$ and $\hat p_i$ are replaced by
$\hat\phi({\svec n})$ and $\hat\pi({\svec n})$ and the position eigenstates
$\hat x_i|x\rangle = x_i|x\rangle$ are replaced by eigenstates
$\hat\phi({\svec n})|\phi\rangle = \phi({\svec n})|\phi\rangle$.  On the
spatial mesh the Hamiltonian density becomes
\begin{eqnarray}
&&\int d^3r\left\{ {1\over 2} \pi^2(r) + {1\over 2} \left|
\nabla\phi(r)\right|^2 + V(\phi)\right\} \nonumber \\
&&\Longrightarrow \sum_{{\svec n}} \left\{ {1\over 2} \pi^2({\svec n}) + {1\over
2} \sum^3_{i=1} \left|\phi ({\svec n} + \mu_i) - \phi ({\svec n}) \right|^2 +
V\left(\phi({\svec n})\right) \right\}
\label{eq:2.11}
\end{eqnarray}
where $\mu_i$ denotes a displacement by one lattice site in the $i^{\rm th}$
direction,
$\sum_{{\svec n}} {1\over 2} \pi^2({\svec n})$
corresponds to the kinetic energy $\sum_j {1\over 2m} \hat p^2_j$, and the
remaining terms, which we will denote as $F[\phi({\svec n})]$ to avoid
confusion with $V[\phi (n)]$ above,
correspond to a sum of one- and two-body potentials
$\sum_{ij} v(\hat x_i,\hat x_j)$.  Introducing
time slices as before yields
\begin{equation}
e^{-\beta\sum_{{\svec n}} \left\{{1\over 2}\pi^2({\svec n}) +
F\left[\phi ({\svec n}) \right] \right\}} = \int {\cal D} \left(\phi_k({\svec
n})\right)\,e^{-\sum_{k,{\svec n}} {1\over 2} \left(\phi_{k+1}({\svec
n}) - \phi_k({\svec n})\right)^2 + F\left[ \phi_k({\svec n})\right]}
\label{eq:2.12}
\end{equation}
The result is a path integral defined on a four-dimensional lattice, for which
we may introduce the obvious notation $n = (n_0, n_1, n_2,n_3)$ where $n_0$
denotes the time label and $n_i$ denotes the spatial label.  One observes that
time slicing replaces $\hat\pi({\svec n})$ by $|\phi_{k+1} ({\svec n}) - \phi_k
({\svec n})|^2 \equiv |\phi(n+\mu_0) - \phi(n)|^2$ which has the same
structure as the discrete spatial derivative $|\nabla\phi|^2 = \sum^3_{i=1}
|\phi(n+\mu_i) - \phi(n)|$.  Hence, a general time-ordered product acquires
the simple form
\begin{equation}
T{\cal O}(\phi) \,e^{-\beta\int d^3r\left\{ {1\over 2}\pi^2 + {1\over
2} (\nabla\phi)^2 + V(\phi)\right\}} \to \int {\cal D}\left(\phi(n)\right) {\cal
O}(\phi)\,e^{-S_{\rm Eucl.} (\phi)}
\label{eq:2.13a}
\end{equation}
where the Euclidean action is
\begin{equation}
S_{\rm Eucl.} (\phi) \equiv \sum_n \left\{ {1\over 2} \sum^3_{i=0}
\left(\phi (n+\mu_i) - \phi(n)\right)^2 + V \left(\phi (n)\right)\right\}
\label{eq:2.13b}
\end{equation}
 
This result merits several comments.  Note that $S_{\rm Eucl.}(\phi)$ is
completely symmetric in space and time, even though the first differences in
space variables arose from a finite-difference approximation to the spatial
derivatives whereas the time differences arose from the path integral time
slicing.  Of course, we are always free to pick different mesh spacings, $a_x$
and $a_t$, in the space and time directions, respectively.  Although in this
derivation, we have gone from $H$ to $S$ using a discrete transfer matrix for
evolution from one time slice to the next, it will often be useful to go
backwards in the other direction to think of the lattice action as describing
evolution of specific states under the Hamiltonians $H$ from one time slice to
another in order to interpret lattice observables.  Depending on the problem,
we may be led to apply different boundary conditions in $x$ and $t$.  In the
case in which all boundary conditions are periodic, the physical problem
corresponds to finite temperature field theory in a periodic three-dimensional
box and the shortest side of the four-dimensional box will effectively act as
the temperature.

\subsection{Coherent States}

We now need to generalize this scalar field result for general second
quantized Fermion or Boson fields.  Recall that the Feynman path integral, and
hence the scalar field path integral, used two basic ingredients:  eigenstates
of $\hat x$, $\hat x|x\rangle = x|x\rangle$, and the resolution of unity, $1 =
\int dx|x\rangle \langle x|$.  The analogs of these relations for creation and
annihilation operators are provided by Boson coherent states.
 
The basic idea is seen most simply for a single creation operator $\hat
a^\dagger$,  corresponding to
a simple harmonic oscillator, for which
\begin{eqnarray}
\left[\hat a,\hat a^\dagger\right]
&=& 1 \nonumber \\
\biggl\{\begin{array}{c}
\hat a^\dagger\\
\hat a
\end{array}
\biggr\} 
|n\rangle 
&=& 
\biggl\{\!
\begin{array}{c}
\sqrt{n+1}\\
\sqrt{n}
\end{array}
\biggr\} 
|n\pm 1\rangle \nonumber \\
|n\rangle &=& {1\over \sqrt{n!}} \left(\hat a^\dagger\right)^n |0\rangle\ \ .
\label{eq:2.14}
\end{eqnarray}
The coherent state $|Z\rangle$ is defined
\begin{equation}
|Z\rangle \equiv e^{Z\hat a^\dagger} |0\rangle = \sum_n {Z^n\over n!}
\left(\hat a^\dagger\right)^n |0\rangle = \sum_n {Z^n\over \sqrt{n!}} |n\rangle
\label{eq:2.15}
\end{equation}
and has the following properties
\begin{eqnarray}
\hat a|Z\rangle &=& \sum_n {Z^n\over \sqrt{n!}} \hat a|n\rangle =
Z\sum_n {Z^{(n-1)}\over \sqrt{(n-1)!}} |n-1\rangle = Z|Z\rangle
\label{eq:2.16a} \\
\langle Z|Z'\rangle &=& \sum_{mn} \bigl\langle m| {Z^{*m}\over
\sqrt{m!}}\;{Z'{}^n\over \sqrt{n!}} |n\bigr\rangle = e^{Z^*Z'}
\label{eq:2.16b} \\
&&\langle Z|: A(\hat a^\dagger,\hat a) :|Z'\rangle = e^{Z^*Z'}\,A(Z^*,Z')
\label{eq:2.16c} \\
&&\int {dZ\,dZ^*\over 2\pi i}
 e^{-Z^*Z} |Z\rangle\langle Z| = 1 
\label{eq:2.16d}
\end{eqnarray}
The last relation is most easily demonstrated by writing the complex variable
in polar form $Z=\rho\,e^{i\phi}$ and performing the $\phi$ integral first.
Analogous results are straightforwardly obtained~\cite{R:JN:10} for a
complete set of creation operators $\hat a^\dagger_\alpha$
\begin{eqnarray}
|Z\rangle
&=& e^{\sum_\alpha Z_\alpha \hat a^\dagger_\alpha}|0\rangle \nonumber \\
\hat a_\alpha |Z\rangle &=& Z_\alpha|Z\rangle  \nonumber \\
\langle Z| :A({\svec a}^\dagger,{\svec a}) :|Z'\rangle &=& e^{\sum_\alpha
Z_\alpha Z'_\alpha} A({\svec Z}^*,{\svec Z})  \nonumber \\
\int \prod_\alpha {dZ^*_\alpha dZ_\alpha\over 2\pi i}\,&& e^{-\sum_\alpha
Z^*_\alpha Z_\alpha} |Z\rangle\langle Z| \equiv\int d\mu(Z) |Z\rangle\langle
Z| = 1 
\label{eq:2.17}
\end{eqnarray}
Proceeding as before, we obtain a path integral by time slicing
\begin{equation}
\bigl\langle Z_f|e^{-\beta H} |Z_i\bigr\rangle = \bigl\langle Z_f
|e^{-\epsilon H} \int d\mu (Z_n) |Z_n\rangle\langle Z_n |e^{-\epsilon H} \int
d\mu (Z_{n-1}) \cdots 
\label{eq:2.18}
\end{equation}
and the matrix element of the infinitesimal evolution operator is
\begin{eqnarray}
d\mu(Z_k)\bigl\langle Z_k| e^{-\epsilon H}| Z_{k-1}\bigr\rangle &=&
\prod_\alpha {dZ^*_{k\alpha} dZ_{k\alpha}\over 2\pi i} e^{-\sum_\alpha
Z^*_{k\alpha}Z_{k\alpha}}\nonumber 
\bigl\langle Z_k|:e^{-\epsilon H(a^\dagger a)}:
+{\cal O}(\epsilon^2)|Z_{k-1}\bigr\rangle \nonumber \\
&=& \prod_\alpha  {dZ^*_{k\alpha}dZ_{k\alpha}\over 2\pi i} e^{-\sum_\alpha
Z^*_{k\alpha}\left(Z_{k\alpha}-Z_{(k-1)\alpha}\right)
-\epsilon H(Z^*_{k\alpha},Z_{(k-1)\alpha})}
\label{eq:2.19}
\end{eqnarray}
with the result
\begin{equation}
\bigl\langle Z_f | e^{-\beta H}|Z_i\bigr\rangle = \int {\cal
D}(Z^*_{k\alpha}, Z_{k\alpha})\,e^{-S(Z^*_{k\alpha},Z_{k\alpha})}
\label{eq:2.20a}
\end{equation}
where
\begin{equation}
S(Z^*,Z) = \sum_k\epsilon\left\{\sum_\alpha Z^*_{k\alpha} \left(
Z_{k\alpha} - Z_{(k-1)\alpha}\right) + H\left(
Z^*_{k,\alpha},Z_{k-1,\alpha}\right)\right\} 
\label{eq:2.20b}
\end{equation}
 
For Fermions with creation and annihilation operators $c^\dagger_\alpha$ and
$c_\alpha$
one must take an additional step and introduce anticommuting
Grassmann variables $\xi$, so that if $\hat c_\alpha|\xi\rangle =
\xi_\alpha|\xi\rangle$ and $\hat c_\beta|\xi\rangle = \xi_\beta|\xi\rangle$,
then we can have
$\hat c_\alpha\hat c_\beta|\xi\rangle = \xi_\alpha\xi_\beta|\xi\rangle = -
\xi_\beta \xi_\alpha|\xi\rangle = - \hat c_\beta \hat c_\alpha|\xi\rangle$.
For our present purposes, one may regard this construction as a set of purely
formal definitions.  Since $\xi^2_\alpha=0$, the only allowable functions are
monomials, functions are specified by the non-vanishing
terms of their Taylor series, and the definite integral is~de\-fined by the properties
$\int d\xi_\alpha=1$ and $\int d\xi_\alpha\xi_\alpha = 1$. Fermion coherent states
are then defined by
\begin{equation}
|\xi\rangle = e^{-\sum \xi_\alpha c^\dagger_\alpha}|0\rangle 
\label{eq:2.21}
\end{equation}
and satisfy relations analogous to (\ref{eq:2.17}) and yield a path integral of the
form (\ref{eq:2.20a}).  Although there are a few technical details which may be
found in
Ref.\cite{R:JN:10}, the essential point is that Grassmann coherent states and
path integrals have essentially the same form as for Bosons, except for a few
crucial minus signs which do all the correct bookkeeping for the difference
between Bosons and Fermions.

\subsection{Gaussian Integrals}

Recall the general formula for the Gaussian integral over complex variables
\begin{equation}
\int \prod_i {dx^*_i\,dx_i\over 2\pi i}\,e^{-x^*_i H_{ij} x_j + J^*_i x_i +
J_i x^*_i} = \left[ \det H\right]^{-1} \,e^{J^*_i H^{-1}_{ij} J_j}
\label{eq:2.22}
\end{equation}
which may be proved by changing to a basis in which $H$ is diagonal and using
$\int dx\,e^{-ax^2} = \sqrt{\pi/a}$.  An
analogous result is obtained for Grassmann variables by noting that
\begin{equation}
\int d\xi^*d\xi\,e^{-\xi^*a\xi} = \int d\xi^*d\xi (1-\xi^*a\xi) = a
\label{eq:2.23}
\end{equation}
Hence
\begin{equation}
\int \prod_i d\xi^*_i d\xi_i\,e^{-\xi^*_i H_{ij} \xi_j + \eta^*_i \xi_i +
\eta_i \xi^*_i} = \left[ \det H\right]\,e^{\eta^*_i H^{-1}_{ij} \eta_j}
\label{eq:2.24}
\end{equation}
and we see that the only difference between complex variables and Grassmann
variables is that $\det H$ appears to the power $-1$ and 1, respectively.
 
With this result, we are prepared to integrate out the Grassmann variables
from the path integral.  Suppose the action has the form
\begin{equation}
S(\xi^*,\xi,\phi) = \xi^*_i M (\phi)_{ij} \xi_j + S_B(\phi) 
\label{eq:2.25}
\end{equation}
where, for example, $\xi^*$, $\xi$ might represent the Fermions
$\bar{\psi}$,  $\psi$ in ${\bar{\psi}(\cslash{p} - \ccslash{A} + m) \psi} +
F_{\mu\nu} (A)^2$ and $\phi$ represents the real Bose field $A$.  Then
\begin{equation}
\int d\xi^*d\xi\,d\phi\,e^{\xi^*M(\phi)\xi + S_\beta (\phi)} = \int
d\phi\, e^{\ln \det M(\phi) + S_\beta(\phi)}
\label{eq:2.26}
\end{equation}
and we are left with an integral over the real field $\phi$ of an effective
action
\begin{equation}
S_{\rm eff} (\phi) = \ln \det M (\phi) + S_B (\phi) 
\label{eq:2.27}
\end{equation}

In the same way, we can perform the Gaussian integrals for propagators.
 Consider first the propagator (or contraction in the language of Wick's
theorem) corresponding to the thermodynamic average of the time-ordered
product of field annihilation and creation operators at space-time points $i =
(x_i t_i)$ and $j = (x_jt_j)$, respectively:
\begin{eqnarray}
\ll T\psi_i\bar{\psi}_j\rr &=& \Tr T\psi_i \bar{\psi}_j\,e^{-\bar{\psi}
M(\hat\phi) \psi + S_B (\hat\phi)} \nonumber \\
&=& \int d\xi^* d\xi\,d\phi\,\xi_i \xi^*_j\,e^{-\xi^* M(\phi) \xi + S_B (\phi)}
\nonumber \\
&=& \int d\phi\,M^{-1} (\phi)_{ij} e^{S_{\rm eff}(\phi)} \ \ .
\label{eq:2.28}
\end{eqnarray}
The last line is obtained by differentiating Eq.~(\ref{eq:2.24}) with respect to
$\eta_i$ and $\eta_j$ which brings down the Grassmann variables $\xi^*_i$ and
$\xi^*_j$ on the left and the inverse matrix on the right.  The general
integral with $n$ pairs of creation and annihilation operators follows
similarly from taking $n$ pairs of derivatives and yields the general form of
Wick's theorem:
\begin{eqnarray} 
\int {\cal D} (\xi^*\xi) \xi_{i_1}&\cdots& \xi_{i_n} \xi^*_{j_n} \cdots
\xi^*_{j_1}\,e^{-\xi^* M\xi} \nonumber \\
&=& {\delta^{2n}\over \delta \eta^*_{i_1}\cdots
\delta \eta^*_{i_n}\delta\eta_{j_n} \cdots \delta
\eta_{j_1}} \int {\cal D}(\xi^*\xi)\,e^{-\xi^*_i M_{ij} \xi_j + \eta^*_i \xi_i +
\eta_i \xi^*_i} \bigg|^{\eta = \eta^* = 0} \nonumber \\
&=& {\delta^{2n}\over \delta \eta^*_{i_1}\cdots
\delta \eta^*_{i_n} \delta \eta_{j_n}
\cdots \delta \eta_{j_1}} \det H\,e^{\eta^*_i M_{ij} \eta_j} \bigg|^{\eta =
\eta^*=0} \nonumber \\
&=& \sum_P (-1)^P M^{-1}_{i_{Pn} j_n} \cdots M^{-1}_{i_{P1} j_1}\,e^{\ln\det
H} 
\label{eq:2.29}
\end{eqnarray}
where $P$ denotes a permutation of the $n$ indices.  Hence, the Fermions may be
integrated out of any physical observable when the action has the form
(\ref{eq:2.25}), leaving the sum of all possible contractions weighted by the effective
action (\ref{eq:2.27}), and we are left with an effective theory containing only
Bosonic degrees of freedom.

\setcounter{equation}{0}
\section{Lattice QCD for Gluons}\label{JWN:sec:3}

It is useful to begin the study of lattice gauge theory with the simplest
possible gauge theory, and gradually increase the generality and complexity
one step at a time.  Hence, in this section we will completely ignore
Fermions, and concentrate only on the pure gluon sector.  This will correspond
to the physical limit in which the quark mass goes to infinity and quarks
cease to be dynamical degrees of freedom.  Furthermore, we will begin with the
simplest possible gauge group, $U(1)$ corresponding to QED, and only after
motivating and displaying the Wilson action for this case will we move to the
non-Abelian $SU(N)$ gauge theory.

\subsection{U(1) Gauge Theory and the Wilson Action}

To motivate the way gauge theory will be formulated on a discrete space-time
lattice, it is useful to recall the essential ideas underlying continuum gauge
theory, and how the entire theory may be viewed as arising from the principle
of gauge invariance.
Therefore, let us consider
a Lagrangian for
a complex scalar field
\begin{equation}
L = \partial_\mu \phi^* \partial_\mu\phi - V(\phi^*\phi) 
\label{eq:3.1}
\end{equation}
and look for  the simplest extension of the theory which is consistent with
local gauge invariance.  Note that since our final goal will be to calculate
Euclidean path integrals, we will always write the Lagrangian and action in
Euclidean form, with the result that $g_{\mu\nu} = \delta_{\mu\nu}$ and upper
and lower Dirac indices are equivalent.
Whereas $L$ is manifestly invariant under the global
gauge transformation $\tilde\phi(x) = -e^{-i\alpha}\phi(x)$, the derivatives in
$L$ yield new terms in the case of a local transformation $\tilde\phi(x) =
e^{-i\alpha(x)} \phi(x)$
\begin{equation}
\partial_\mu \tilde\phi^* \partial_\mu \tilde\phi = \left[ \left(
\partial_\mu - i (\partial_\mu\alpha)\right)\phi\right]^* \left(\partial_\mu -
i (\partial_\mu\alpha)\right)\phi
\label{eq:3.2}
\end{equation}
If we adopt the principle of local gauge invariance, that the theory should be
independent of the arbitrary phase choice $\alpha(x)$ that various observers
might choose at different points in space, then we may repair the theory by
adding a ``compensating'' field $A_\mu(x)$ such that
\begin{equation}
\tilde A_\mu (x) = A_\mu (x) +{1\over g} \partial_\mu \alpha(x)
\label{eq:3.3}
\end{equation}
If we now replace the derivative $\partial_\mu$ in $L$ by the covariant
derivative
\begin{equation}
D_\mu\phi(x) \equiv \left(\partial_\mu + i gA_\mu (x)\right)\phi(x)
\label{eq:3.4}
\end{equation}
we observe that the transformation of $A_\mu(x)$ exactly compensates for the
undesired derivative of $\alpha$ and yields an invariant Lagrangian.
 
Whereas the coupling of the new field $A$ to $\phi$ was determined from gauge
invariance, the only guiding principle for determining the action for $A$
itself is simplicity and economy.  Thus, we seek  the simplest action
for $A_\mu$ involving the least number of derivatives
which is consistent with gauge invariance and Lorentz invariance.
Noting that
\begin{equation}
\partial_\mu\Bigl(A_\nu +{1\over g}
\partial_\nu\alpha\Bigr)  - \partial_\nu \Bigl(A_\mu +{1\over g}
\partial_\mu\alpha\Bigr) = \partial_\mu A_\nu - \partial_\mu A_\mu
\equiv F_{\mu\nu}
\label{eq:3.5}
\end{equation}
we observe that $F_{\mu\nu}$ is gauge invariant so that $F_{\mu\nu}{}^2$ is both
gauge and Lorentz invariant and we are thus led automatically to Maxwell's
equations and the complete Lagrangian
\begin{equation}
L = - {1\over 4e^2} \left(\partial_\mu A_\nu - \partial_\nu A_\mu\right)^2 +
\left(D_\mu\phi\right)^* \left( D_\mu\phi\right) - V\left(\phi^*\phi\right)
\ \ .
\label{eq:3.6}
\end{equation}
For subsequent treatment on a lattice, it is useful to note that the
appropriate operator to compare fields at two different points $x$ and $y$ is
the link variable
\begin{equation}
U(y,x) = e^{i\int^y_x dx_\mu g A_\mu (x)} 
\label{eq:3.7}
\end{equation}
which simply removes the arbitrary phases between the two points and yields a
gauge invariant result.
 
We now consider how to approximate this continuum theory on a space-time
lattice.  Often in numerical analysis, one may allow discrete approximations
to break fundamental underlying symmetries.  For example, when one solves the
time-independent Schr\"odinger equation on a spatial mesh, one violates
translational invariance.  There may be small spurious pinning forces which
reflect the fact that the energy is slightly lower when the solution is
centered on a mesh site or centered between mesh sites, but there are no major
qualitative errors and the quantitative errors may be strictly controlled.
When one solves the time-dependent Schr\"odinger equation or time-dependent
Hartree--Fock equation, however, one finds that it is important to enforce
certain properties such as energy conservation and unitarity when discretizing the
problem in time.  In the case of lattice gauge theory, since by the previous
argument gauge invariance plays such a crucial role in defining the theory, it
is desirable to enforce it exactly in the lattice action.  In contrast, as in
the case of the Schr\"odinger equation, we will settle for an action which
breaks Lorentz invariance, and simply insist on making the lattice spacing
small enough that the errors are acceptably small.
 
Following Wilson, we define the action in terms of directed link variables
assigned to each of the links between sites of the space-time lattice.  For
$U(1)$, we define the link variable from site $n$ in the $\mu$ direction to
site $n+\mu$ as a discrete approximation to the integral $e^{ig\int^{n+\mu}_{n}
dx\,A_\mu}$ which we denote
\begin{equation}
U_\mu (n) = e^{i\theta_\mu(n)} = e^{iag A_\mu (n+{a \over 2} \mu)}\ \ .
\label{eq:3.8}
\end{equation}

Thus $\theta_\mu(n)$ is a discrete approximation to $g\int^{n+\mu}_n dx\,A_\mu$
along the direction of the link, with $A_\mu$ evaluated at the center of the link. 
When the direction is reversed,
$U_n(n)\to U_n(n)^\dagger$.  The link
variable is then a group element of $U(1)$ and the compact variable
$\theta_\mu(n)$ will be associated with $agA_\mu (x)$ in the continuum limit.
With these link variables, the integral over the field variables in the path
integral is replaced by the invariant group measure for $U(1)$, which is
${1\over 2\pi}\int^\pi_{-\pi} d\theta$.
 
\begin{figure}
$$\BoxedEPSF{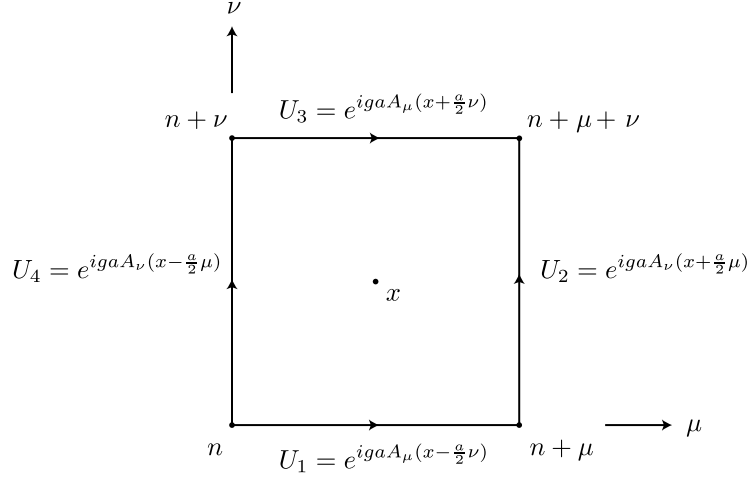}$$
\caption{An elementary plaquette of link
variables.}\label{F:JNFig1}
\end{figure}

The fundamental building blocks of the lattice action are products of directed
link variables taken counter-clockwise around each individual plaquette of the
lattice.  By construction, this product is gauge invariant, ensuring gauge
invariance of the resulting action.  A typical plaquette is sketched in
Fig.~\ref{F:JNFig1},  $\mu$ and $\nu$ are unit vectors in the horizontal and vertical
directions,  the position of the center of the plaquette is $x$, and we express the centers of
each link in terms of displacements $\pm {a \over 2}$ in the $\mu$ or $\nu$ directions. 
The product of the four group elements around the plaquette may thus be written
\begin{eqnarray}
U^{\sq}_{\mu \nu} &=& \prod_{\sq} U_1 \,U_2 \,U^\dagger_3 U^\dagger_4 
\nonumber
\\ 
&=& e^{iag [A_\mu (x-{a \over 2} \nu) + A_\nu (x+{a \over 2} \mu)
-A_\mu (x+{a \over 2} \nu) - A_\nu (x-{a \over 2} \mu)]} \nonumber \\
&\cong& e^{ia^2g (\partial_\mu A_\nu - \partial_\nu A_\mu)} \nonumber \\
&=& e^{ia^2g F_{\mu\nu}} \ \ .
\label{eq:3.9}
\end{eqnarray}
Note that the discrete lattice difference operator $A_\nu (x+{a \over 2} \mu)
-A_\nu (x-{a \over 2} \mu)$
becomes the derivative $a \partial_\mu A_\nu$ in the continuum limit, so that
the exponent is a discrete approximation to the curl on the lattice and is
proportional to
$\partial_\mu A_\nu - \partial_\nu A_\mu = F_{\mu\nu}$ in the continuum.
Since each plaquette generates an approximation to $F_{\mu\nu}$, an action
which corresponds to QED in the continuum limit may be constructed by choosing
a function of $U^{\sq}_{\mu \nu}$ which yields $F^2_{\mu\nu}$ plus terms that
are negligible in the continuum limit.  Defining the inverse coupling constant $\beta_g = {1
/ g^2}$, where the subscript~$g$ distinguishes it from other quantities commonly denoted
by~$\beta$, and expanding in the limit $a\to 0$, the action may be written
\begin{eqnarray}
S(U) &=& \beta_g \sum_{\sq} \left( 1 - \Re U^{\sq}\right) \nonumber  \\
 &\sim& {1\over g^2} \sum_{\sq} \left( 1 - \cos\left( a^2 g
F_{\mu\nu}\right)\right) \nonumber \\
&\sim& {1\over g^2} \sum_{n\{\mu\nu\}} \Bigl( {a^4 g^2\over 2} F^2_{\mu\nu} (n)+
\cdots\Bigr) \nonumber \\
&\sim& {1\over 2} \sum_n a^4 \sum_{\{\mu\nu\}} F^2_{\mu\nu}(n) \nonumber \\
&&\to {1\over 4} \int
d^4x F_{\mu\nu}(x)F_{\mu\nu} (x)\ \ .
\label{eq:3.10}
\end{eqnarray}
In the second line $\{\mu\nu\}$ denotes the sum over all pairs of $\mu$ and
$\nu$ arising from the sum over plaquettes and the extra factor of 1/2 in the
last line accounts for the fact that each pair occurs twice in the double sum
over repeated indices $F_{\mu\nu}F_{\mu\nu}$.  The terms
higher order in the lattice cutoff $a$ vanish in the classical
continuum limit and may give rise to finite renormalization of the coupling
constant in quantum field theory.
The lattice gauge theory defined by (\ref{eq:3.10}) is in a form which may be solved
directly using the Metropolis or heat bath
methods.  For readers worried that we have not systematically accounted for all
corrections of higher order in $a$, a more complete derivation will be presented later
for the general $SU(N)$ case.
 
It is useful to examine the role of Gauss' law and how
the presence of external charges is manifested in this theory.  The basic ideas
are most easily sketched in the continuum theory.  Since the Hamiltonian does
not constrain the charge state of the system, we must project the states
appearing in the path integral onto the space satisfying ${\svec\nabla}\cdot
{\svec E} = \rho$ with a specific background charge $\rho$, which may be
accomplished by writing a $\delta$-function in the form $\int {\cal D}
\chi\,e^{i\int dx\,dt\,\chi( {\svec\nabla}\cdot {\svec E}-\rho)}$.
Remaining in temporal gauge $A_0=0$ and using the form of the path integral
(2.2) in which both the
coordinate $x\to A$ and momentum $\rho\to E$ appear, the
path integral for the partition function projected onto the space with
external source $\rho$ may be written
\begin{eqnarray}
Z &=& \int {\cal D}\chi\,{\cal D}{\svec A} \,{\cal D}{\svec
E}\,e^{\int dx\,dt \bigl[ i {\svec E}\cdot\dot{\svec A} - {1\over 2}
 (E^2 + B^2) + i \chi ( {\svec\nabla}\cdot{\svec E} -
\rho)\bigr]} \nonumber \\
&=& \int {\cal D}\chi\,{\cal D}{\svec A}\, e^{- \int dx\,dt\bigl\{ {1\over 2}
\bigl[ \bigl( \dot{\svec A} - {\svec\nabla}\chi\bigr)^2 + B^2 - i \chi
\rho\bigr]\bigr\}}\ \ .
\label{eq:3.16a}
\end{eqnarray}
Equation (\ref{eq:3.16a}) is an important result.  Having started in temporal gauge
$A_0=0$, we see that enforcing Gauss' law gives rise to a projection integral
over an additional field $\chi$ which enters into the final action just like
the original $A_0$ field.  Indeed, renaming $\chi = A_0$ so that $\dot A_i -
\partial_i A_0 = F_{0i}$ and writing the source as a set of point charges
$\rho(x) = \sum_n q_n \delta(x-x_n)$, we obtain
\begin{equation}
Z = \int {\cal D} A_\mu \,e^{-\int dx\,dt {1\over 4} F_{\mu\nu}
F_{\mu\nu}}
\prod_n e^{-iq_n\int dt A_0 (x_n,t)}\ \ .
\label{eq:3.16b}
\end{equation}
Thus, the Hamiltonian path integral with projection is precisely the
Lagrangian path integral with a line of $\pm A_0$ fields at the positions of
the fixed external $\pm$ charges.  In the case of no external charges, we may
think of the Lagrangian path integral including the $A_0$ integral as the
usual filter $e^{-\beta H}$ selecting out the ground state.  In the
presence of charges, the path integral augmented by lines of $A_0$ at the
positions of the charges filters out the ground state in the presence of these
sources.

\subsection{SU(N) Gauge Theory}

The generalization to non-Abelian gauge theory is straightforward.  The link
variables become group elements of $SU(N)$
\begin{equation}
U_\mu(n) = e^{i\,ag{1\over 2} \lambda^c A^c_\mu(n)} \equiv
e^{i\,ag\tilde A_\mu(n)}
\label{eq:3.17}
\end{equation}
where the $\lambda^c$ are Pauli matrices or Gell--Mann matrices for $SU(2)$ or
$SU(3)$ and $c$ is a color label which runs over the $N^2-1$ generators
$\lambda^c$.   The integration in the path integral is defined by the
invariant group measure which we will denote by ${\cal D}(U)$.

Since derivation of the $SU(N)$ action is one of our primary results, let us be slightly
more careful than in the $U(N)$ case and keep track of higher order terms in $a$ to
ascertain the leading error in our result.\cite{M:JN:12}  As before, we refer the
centers of each of the links in Fig.~1 to the point $x$ in the center of the plaquette,
so that for
$SU(N)$
\begin{eqnarray}
U_1 &=& e^{iga \tilde{A}_\mu (x-{a \over 2}\nu)} \qquad \qquad 
U_2 = e^{iga \tilde{A}_\nu (x+{a \over 2}\mu)} \nonumber \\
U_3 &=& e^{iga \tilde{A}_\mu (x+{a \over 2}\nu)} \qquad \qquad 
U_4 = e^{iga \tilde{A}_\nu (x-{a \over 2}\mu)}
\label{eq:3.18}
\end{eqnarray}
and the product of $SU(N)$ group elements around an elementary plaquette is
\begin{equation}
U^{\sq}_{\mu \nu} = \prod_{\sq} U_1 \,U_2 \,U^\dagger_3 \,U^\dagger_4
\label{eq:3.19a}
\end{equation}
With this notation, we note that $U_1(-a) = U^\dagger_3 (a)$ and $U_2(-a) = U^\dagger_4
(\dot{a})$.  Hence, $ \Tr U^{\sq}_{\mu \nu}  (-a) = \Tr U^\dagger_3 U^\dagger_4 U_1 U_2
= \Tr U^{\sq}_{\mu\nu} (a)$ so that $ \Tr U^{\sq}_{\mu \nu} $ is an even function
of $a$.   Also, because $U^{\sq}_{\mu \nu}$ is a product of unitarity matrices,
$U^{\sq}_{\mu \nu} U^{\sq\dagger}_{\mu \nu} = 1$.  The continuum contribution
is obtained by expanding $\tilde{A}_\mu (x -{a \over 2} \nu) = \tilde{A}_\mu (x) - {a
\over 2} \, {\partial \over \partial \nu} \tilde{A}_\mu + \theta (a^2)$ as before and
applying the Baker--Hausdorff identity
$
e^{\hat{x}} e^{\hat{y}} = e^{(\hat{x}\hat{y}+ {1 \over 2} [\hat{x}\hat{y}] + \cdots)}
$
to each quantity below in curly brackets, with the result
\begin{eqnarray}
U^{\sq}_{\mu \nu} &=& \Bigl\{ e^{iag (A_\mu -{a \over 2} \partial_\nu A_\mu +
{\cal O} (a^2))} e^{ig  (A_\nu +{a \over 2} \partial_\mu A_\nu + {\cal O} (a^2))}
\Bigr\} 
 \nonumber \\
 && {} \times  
\Bigl\{ e^{iag (A_\mu +{a \over 2} \partial_\nu A_\mu +
{\cal O} (a^2))} e^{ig  (A_\nu -{a \over 2} \partial_\mu A_\nu + {\cal O} (a^2))} \Bigr\} 
\nonumber \\
&=&  e^{iag (A_\mu + A_\nu +{a \over 2} (\partial_\mu A_\nu - \partial_\nu
A_\mu) + {iag \over 2} [A_\mu, A_\nu] + {\cal O} (a^2))} 
 \nonumber \\
  &&{} \times
e^{iag (-A_\mu - A_\nu +{a \over 2} (\partial_\mu A_\nu - \partial_\nu
A_\mu) + {iag \over 2} [A_\mu, A_\nu] + {\cal O} (a^2))} 
\nonumber \\
&=& e^{iag[a (\partial_\mu A_\nu - \partial_\nu A_\mu) + iag [A_\mu, A_\nu] +
{\cal O} (a^2)]}
\label{eq:3.19b}
\end{eqnarray}
Writing $F_{\mu\nu} \equiv \partial_\mu A_\nu - \partial_\nu A_\mu + iag [A_\mu,
A_\nu] $ and defining the next two terms in an expansion in powers of $a$ as
$G_{\mu\nu}$ and $H_{\mu\nu}$, we obtain
\begin{eqnarray}
U^{\sq}_{\mu \nu} &=& e^{ia^2 g F_{\mu\nu} + ia^3G_{\mu\nu} + ia^4
H_{\mu\nu}} 
\nonumber \\
&=& 1+ ia^2 F_{\mu\nu} + ia^3G_{\mu\nu} + ia^4 H_{\mu\nu} - {a^4 \over 2} g^2
F_{\mu\nu}^2 + \theta(a^5)
\label{eq:3.19c}
\end{eqnarray}
Unitarity of $U^{\sq}_{\mu \nu}$ implies that $F$, $G$ and $H$ are Hermitian
which combined with the fact that $\Tr U^{\sq}_{\mu \nu}$ is an even function
of $a$ yields
\begin{equation}
\Re \Tr U^{\sq}_{\mu \nu} = \fracs12 \Tr (U^{\sq}_{\mu \nu} +
U^{\sq\dagger}_{\mu \nu}) = \Tr (1 -  \fracs12 a^4 g^2 F_{\mu\nu}^2 ) + \theta(a^6)
\label{eq:3.19d}
\end{equation}

Defining the inverse coupling $\beta_g=2N/g^2$, the $SU(N)$ action may be written
\begin{eqnarray}
S(U) &=& \beta_g \sum_{\sq} \Bigl(1-\frac1N  \Re \Tr U_{\sq}\Bigr) \nonumber \\
 &=& \frac{\beta_ga^4g^2}{2N} \sum_{n\{\mu\nu\}} \Tr(\fracs12 \lambda^c F_{\mu\nu}^c
\fracs12\lambda^b F_{\mu\nu}^b) + {\cal O}(a^6) \nonumber\\
&=& \sum_n a^4 \sum_{\{\mu\nu\}} \fracs12  F_{\mu\nu}^c (n) F_{\mu\nu}^c (n)  
+ {\cal O}(a^6) \nonumber\\
&&  \to \fracs14 \int d^4x\, F_{\mu\nu}^c (x) F_{\mu\nu}^c (x) 
+ {\cal O}(a^6)\ \ . 
\label{eq:3.15}
\end{eqnarray}
Note that summation over repeated indices is implied everywhere except where
$\sum_{\{\mu\nu\}}$ denotes sums over distinct pairs, and the property $\Tr
\lambda^b\lambda^c = 2\delta_{bc}$ has been used in the second line.

\subsection{Wilson Loops and Lines}

Rephrasing the original discussion of gauge fields in terms of lattice
variables, if there were a quark field defined on a lattice, then under a
local gauge transformation, the field $\psi_i$ at each site would be
multiplied by a group element $g_i$.  The link variables were explicitly
introduced to compensate such a gauge transformation, so the link variable
$U_{ij}$ going from site $i$ to $j$ is multiplied by $g_i$ and $g^{-1}_j$.
Thus, the overall effect of a gauge transformation is the following:
\bea
U_{ij} &\to& g_i U_{ij} g^{-1}_j \nonumber\\
\psi_i &\to& g_i \psi_i  \nonumber\\
\bar{\psi}_i &\to& \bar{\psi}_i g^{-1}_i \ \ . 
\label{eq:3.25}
\eea
 
In the pure gauge sector, where the only variables are link variables, it is
clear from (\ref{eq:3.25}) that the only gauge invariant objects which can be
constructed are products of link variables around closed paths, for which the
factors of $g$ and $g^{-1}$ combine at each site.  The Wilson loop is
therefore defined as the trace of a closed loop of link variables
\bea
W &\equiv& \Tr \prod\limits_{ij\in c} U_{ij} \nonumber\\
    &=& \Tr U_{ij} U_{jk} \cdots U_{mn} U_{ni} \label{eq:3.26}
 \eea
and specifies the rotation in color space that a quark would accumulate along
the loop $c$ from the path-ordered product $P_c\,e^{\int ig\tilde A}$.
 
To understand the physical significance of a space-time Wilson loop, it is
useful to note that by (\ref{eq:3.25}), a quark creation operator at site $j$ and an
annihilation operator at site $i$ transform under gauge transformations like a
product of link variables connecting sites $i$ and $j$
\bea
\psi_i \bar{\psi}_j &\to& g_i \psi_i \bar{\psi}_j g^{-1}_j \nonumber\\
U_{ik} U_{k\ell}\cdots U_{mj} &\to& g_i U_{ik} U_{k\ell}\cdots U_{mj} g^{-1}_j\ \
. \label{eq:3.27}
\eea
Thus, as far as the gluon fields are concerned, the ends of a chain of link
variables are equivalent to an external quark-antiquark source, and the
presence of such a chain of link variables therefore measures the response of
the gluon fields to an external quark-antiquark source.
\begin{figure}
$$\BoxedEPSF{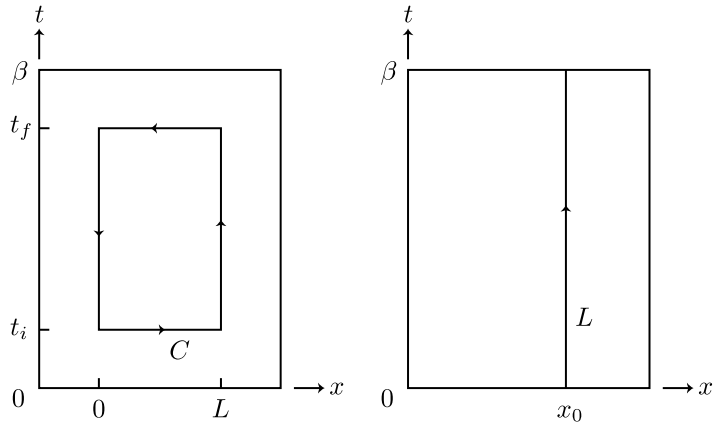}$$
\caption{A space-time Wilson loop defined by the
chain of link variables $C$ on a finite lattice (left) and a Wilson or
Polyakov line defined by the chain of link variables $L$ on a finite lattice
(right).}
\label{JN:Fig2}
\end{figure}%
 
Now, consider the time evolution of the system corresponding to the
expectation value of the Wilson loop drawn in Figure~\ref{JN:Fig2}
\be
\ll W\rr = {\int dU\,e^{-S(U) }\Tr \prod\limits_C U_{ij} \over \int
dU\,e^{-S(U)}} \ \ .\label{eq:3.28}
\ee
Prior to the time $t_i$, there are no color sources present, so evolution
filters out the gluon ground state in the zero charge sector, $|0\rangle =
e^{-t_iH}$ $|Q = 0\rangle$.  At time $t_i$, the line of link variables
between 0 and $L$ creates an external antiquark source at 0 and a quark source
at $L$.  As discussed in connection with Eq.~(\ref{eq:3.16b}), the links in the time
direction between $t_i$ and $t_f$ maintain these sources at 0 and $L$.  Hence,
for any $t$ between $t_i$ and $t_f$, the evolution filters out the lowest
gluon configuration in the presence of external quark-antiquark sources
producing the state
$|\psi\rangle = e^{-(t-t_i)H} \psi(0) \bar{\psi}(L)|0\rangle$.  Finally, at
time $t_f$, the external sources at 0 and $L$ are removed by a line of links
from $L$ to 0, and the system is returned to the zero charge sector.  Using
Feynman's picturesque language of antiquarks corresponding to quarks
propagating backwards in time, one may succinctly characterize the Wilson loop
as measuring the response of the gluon fields to an external quark-like
source traveling around the perimeter of the space-time
loop in the direction of the arrows.
 
Quantitatively, if $t_f- t_i$ is large enough, the lowest gluon state in the
presence of quark and antiquark sources separated by $L$ will dominate, and
$\ll W\rr$ will be proportional to $e^{-(t_f-t_i) V(L)}$ where $V(L)$ is the
static quark-antiquark potential. Physically, this potential corresponds to
the potential arising in heavy quark spectroscopy.  Furthermore, at large distances,
the potential in the pure gluon sector becomes linear (since the flux tube cannot
be broken by $q\bar{q}$ pair creation) so the Wilson loop enables direct
numerical calculation of the string tension.  If the Wilson loop has $I$ links in the
time direction and $J$ links in the space direction, then
\bea
W(IJ) = \Bigl\langle \Tr \prod\limits_{C_{IJ}}U \Bigr\rangle
&\mathop{\sim}\limits_{I\to\infty}& e^{-aI V(aJ)} \nonumber\\
 &\mathop{\sim}\limits_{I,J\to\infty}& e^{-a^2 \sigma IJ} \ \ .
\label{eq:3.29}
\eea
 
The exponent is thus proportional to the area for large loops, and this area
behavior is a signature of confinement, since it arises directly from the
linearly rising potential.  Although the preceding physical argument was
framed in Hamiltonian form with evolution in the time direction, it is clear
that because of the symmetry of the Euclidean action, all space-time
dimensions are equivalent and the area law reflects this symmetry.

Since the string tension can be calculated directly from Wilson loops, it
is useful to relate it to an experimentally measurable quantity, the slope of
Regge trajectories.  It is an empirical fact that families of meson states
with a given set of internal quantum numbers have mass dependence on the total
angular momentum $J$ which is accurately described by the Regge formula
\be
M^2_J = {1\over \alpha'} J\ \ ,\qquad \alpha' = 0.9\,\hbox{GeV}^{-2}\ \
.\label{eq:3.30}
\ee
To see how the slope $\alpha'$ is related to the string tension, it is useful
to consider a very simple model in which a massless quark and antiquark are
connected by a string or flux tube of length $2L$.
Since the quarks are massless, they must move at the speed of light, and the
velocity of the segment of string a distance $x$ from the origin is $v={x\over
L}$.  If $\sigma$ is defined as the energy per unit length of the flux tube in
its rest frame, then the contributions of the element of length $dx$ at point
$x$ to the energy and angular momentum are
\be
dE = \gamma\,\sigma\, dx\ \ ,\qquad
dJ = \gamma\sigma\,vx\,dx \label{eq:3.31a}
\ee
where $v={x\over L}$ (with $c=1$) and $\gamma=\left(1-v^2\right)^{-1/2}$.
Hence,
\be
M = \int^L_{-L} {\sigma\,dx\over \sqrt{ 1-\left( {x\over
L}\right)^2}} = \pi\sigma L \ \ ,\quad
J = \int^L_{-L} {\sigma{x^2\over L}dx\over \sqrt{ 1-\left( {x\over
L}\right)^2}}= {\pi\over 2} \sigma L^2 
\label{eq:3.31b}
\ee
and
\be
M^2 = \pi^2 \sigma^2 L^2 = 2\pi\sigma J \label{eq:3.31c}
\ee
so that
\be
\sqrt{\sigma} = \left[ 2\pi \alpha'\right]^{-1/2} = 420\,\hbox{MeV}\ \ .
\label{eq:3.32}
\ee
Clearly this flux tube model is a drastic oversimplification, especially for
low angular momentum states for which the finite width of the tube, the
structure of the end caps, and the lack of localization of the quarks could
all produce large corrections.  At large angular momentum, however,
the picture is
somewhat more convincing.  In the context of this model, I will therefore take
the point of view that the accurate Regge behavior at low angular momentum is
accidental and that $\sigma$ is primarily determined by high angular
momentum states.  However, since the ultimate theory is undoubtedly much more
complicated, I will not regard it as a serious problem if lattice parameters
determined from this rough argument disagree at the 5~to 10\% level relative
to those determined from other observables such as direct calculation of
hadron masses.
 
Although the area law behavior of large Wilson loops is clear from (\ref{eq:3.29}), in
practical calculations on finite lattices, there are significant corrections,
including a term proportional to the perimeter arising from the self-energy of
the external sources and a constant arising from gluon exchanges at the
corners, so that
\be
 - \ln W(I,J) \sim C+D (I+J) + a^2 \sigma IJ\ \ . \label{eq:3.33}
\ee
To eliminate the constant and perimeter terms, the following ratio  introduced by
Creutz \cite{R:JN:06} of Wilson
loops having the same perimeter is calculated to cancel out the $C$ and $D$
terms in (\ref{eq:3.33})
\be
\chi(I,J) = - \ln \left( {W(I,J) W(I-1, J-1)\over W(I,J-1) W(I-1, J)}\right)
\sim a^2\sigma\ \ .
\label{eq:3.34}
\ee
 
A Wilson or Polyakov line is another form of gauge invariant closed loop which
can be placed on a periodic lattice.  In this case, as sketched in Fig.~2, the
links are located at a fixed position in space $x_0$ and run in the time
direction from the first time slice to the last, which by periodicity is
equivalent to the first and thus renders the product gauge invariant.  If the
length of the lattice in the time direction is $\beta_t$ (were the subscript $t$
distinguishes it from the inverse coupling constant $\beta_g \equiv {2N\over
g^2}$), the expectation
value of the Wilson line yields the partition function for the gluon field in
the presence of a single fixed quark at inverse temperature $\beta_t$ and thus
specifies the free energy $F_{\rm quark}$ of a single quark 
\bea
\bigl\langle\hat L\bigr\rangle &=& {\int {\cal D}U\,e^{-S(U) } \Tr \prod_L U_{ij}
\over \int {\cal D} U\,e^{-S(U)}} \nonumber\\
&=& e^{-\beta_t F_{\rm quark}} \ \ .
\label{eq:3.35}
\eea
This quantity is useful as an order parameter for the
deconfinement phase transition. Note that because the periodic lattice is a
four-dimensional torus and $L$ winds around the lattice once in the time
direction, it is characterized by a winding number and is thus topologically
distinct from a Wilson loop which has winding number 0.  By the preceding
argument, two lines in opposite directions, one at $x=0$ and one at $x=L$, will
produce the free energy of a quark and antiquark separated by distance $L$,
and as $\beta_t\to\infty$ this provides a means of calculating
the average of the singlet and octet static quark-antiquark potentials.

\subsection{Strong Coupling Expansion}
One can obtain a useful physical picture of what happens when one evaluates
the expectation value of a Wilson loop
\be
\ll W\rr = Z^{-1} \int {\cal D}(U) \,e^{-\beta_g\sum_{\sq} \left( 1 -
{1\over 2N} \tr\left( U_{\sq} + U^\dagger_{\sq}
\right)\right)} \tr \prod_C
U \label{eq:3.36}
\ee
by expanding the exponent in powers of the inverse coupling constant $\beta_g
\equiv {2N\over g^2}$.  Since this expansion requires small $\beta_g$ and thus
large $g^2$, it is called the strong coupling expansion.  Note that although
it is formally analogous to the high temperature expansion in statistical
mechanics, where $\beta_g$ would be replaced by the inverse temperature, the
actual physical temperature of our system is specified by the length of the
lattice in the time direction, $\beta_t$, and is distinct from $\beta_g$.
 
The structure of the expansion is revealed by considering the integrals over
group elements which arise in the path integral (\ref{eq:3.36}).  A general discussion
of integration over $SU(N)$ group elements is given by Creutz~\cite{R:JN:06}, but for
our present purposes it is sufficient to use the following
two results, where Greek indices
denote $SU(N)$ matrix indices, not sites
\bea
\int dU\,U_{\alpha\beta} &=& 0 \label{eq:3.37a}\\
\int dU\,U_{\alpha\beta} U^{-1}_{\gamma\delta} &=& {1\over
N}\delta_{\alpha\delta} \delta_{\beta\gamma}
\label{eq:3.37b}
\eea
These results follow directly from the orthogonality relation for irreducible matrix
representations of the group and are trivially verified for $U(1)$ for which
the invariant measure is $\int dU = {1\over 2\pi} \int^\pi_{-\pi}d\theta$ and
$U = e^{i\theta}$.
\begin{figure}
$$\BoxedEPSF{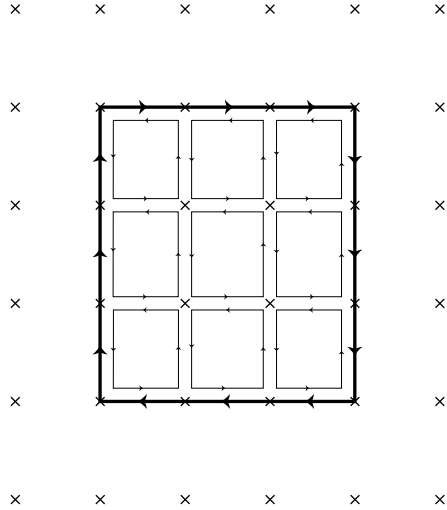}$$
\caption{A $3\times 3$ Wilson loop tiled with plaquettes
in the strong coupling expansion.}
\label{JN:Fig3}
\end{figure}
 
Now consider the diagrams which result from drawing the links in the Wilson
loop $\prod_C U$ and some set of plaquettes $U_{\sq}$ and
$U^\dagger_{\sq}$
obtained from expanding the exponential in (\ref{eq:3.36}).  The integral (\ref{eq:3.37a})
tells us that any diagram which has a single exposed link (that is, a single link
between a pair of sites) anywhere on the lattice gives no contribution.  Thus,
the only non-vanishing terms in the expansion are those in which we manage to
mate plaquettes from the exponential with the Wilson loop to eliminate all
exposed links.  The simplest way to mate two links to obtain a non-vanishing
result is to place them between the same sites in opposite directions, which
by (\ref{eq:3.37b}) yields ${1\over N}$.
 Since each plaquette brings with it a factor of $\beta$, the lowest order
non-vanishing contribution to $\ll W\rr$ is obtained by ``tiling'' the
interior of the Wilson loop with plaquettes oriented in the opposite direction
as sketched in Fig.~\ref{JN:Fig3} for a $3\times 3$ loop.  Note that each of the outer
links of the original Wilson loop is protected, and the interior links protect
each other pairwise.  The leading term for an $I\times J$ Wilson loop would
thus have $I\times J$ tiles, each contributing a factor ${\beta\over 2n}$.  In
addition, because of the traces in the plaquettes and Wilson loop in (\ref{eq:3.36})
and the $\delta$'s in Eq.~(\ref{eq:3.37b}) there is a factor of $N$ for each of the
$(I+1)(J+1)$ sites and because of the factor ${1\over N}$ in (\ref{eq:3.37b}) there is
a factor ${1\over N}$ for each of the $(2IJ+I+J)$ double bonds.  Hence, except
for $SU(2)$, where the counting is different because the two orientations of
plaquettes are equivalent, the overall contribution goes as
\be
W(IJ) \sim \Bigl( {\beta\over 2N^2}\Bigr)^{IJ} \label{eq:3.38a}
\ee
giving the lowest order contribution to the string tension
\be
\sigma\sim - a^{-2} \ln \Bigl( {\beta\over 2N^2}\Bigr)\ \ .
\label{eq:3.38b}
\ee
Fancier tilings are also possible if one is willing to use more tiles and thus
include more powers of $\beta$.  For example, one could place five tiles
together to make a cubic box with an open bottom and replace one or more tiles
with this box.  The box could be elongated, or even grown into a tube which
connects back somewhere else. Alternatively, one could replace a plaquette
oriented in one direction by $(N-1)$ plaquettes oriented in the opposite
direction to obtain a non-vanishing $SU(N)$ integral.
 
The utility of this
expansion is twofold.
It provides a physical picture of filling in the
Wilson loop with a gluon membrane, whose vibrations and contortions represent
all the quantum fluctuations of the gluon field.  When observed on a
particular time slice, the cross section of this surface corresponds to a
color flux tube joining the quark-antiquark sources.
In addition, in low orders, the individual terms can be calculated
explicitly and provide a valuable quantitative check of numerical
calculations.

\subsection{Continuum Limit and Renormalization}
Pure gauge theory on a finite lattice is specified by two parameters: the
dimensionless bare coupling constant $g$ and the lattice spacing $a$
corresponding to a momentum cutoff $p_{\rm max}\sim {\pi\over a}$.  As $a$ is
changed, the bare $g$ must be changed to keep physical quantities fixed.
 
In principle, the renormalization procedure on a lattice is very simple and
could be carried out as follows.  First, pick an initial value of $g$ and
calculate some set of dimensionful physical observables $\ll {\cal O}_i\rr$.
These observables may be written in the form
\be
\ll {\cal O}_i\rr = a^{-d_i}\ll f_i(g)\rr \label{eq:3.46}
\ee
where $d_i$ is the dimension of the operator and $f_i$ is the dimensionless
quantity calculated on the lattice using the Wilson action with $\beta =
{2N\over g^2}$ and with all lengths expressed in units of the lattice spacing
$a$.  For example, we have already seen in Eq.~(\ref{eq:3.34}) that the string tension
has the form $\sigma = a^{-2}\chi$.  Then, use the physical value of one
operator, say ${\cal O}_1$, to determine the physical value of $a$
corresponding to the selected $g$.  Again, using the string tension example,
we could define $a = {\sqrt{\chi}/ 420\,\hbox{MeV}}$.  With this value of
$a$ determined from ${\cal O}_1$, all other observables ${\cal O}_2\cdots {\cal
O}_N$ are completely specified.  One should then repeat this procedure for a
sequence of successively smaller and smaller values of $g$, thereby
determining the function $a(g)$ and a sequence of values for the observables
${\cal O}_2\cdots {\cal O}_N$.  If the theory is correct, then each sequence
of observables ${\cal O}_i$\ $i\not=1$ should approach a limit as $g\to 0$,
and that limit should agree with nature.
 
In practice, it would be very difficult to carry out a series of calculations
as described above to small enough $g$ to make a convincing case.  Hence, it
is preferable to make use of our knowledge of the relation between the
coupling constant and cutoff based on the renormalization group in the
perturbative regime, and only carry out explicit lattice calculations down to
the point at which the renormalization group behavior is clearly established.
The foundation of the argument is the fact that the first two coefficients in
the expansion of the renormalization group function $a{dg\over da}$ are
independent of the regularization scheme, and thus may be taken from continuum
one and two loop calculations:
\be
a {dg\over da} = \beta_0 g^3 + \beta_1 g^5 + \cdots
\label{eq:3.47}
\ee
where $\beta_0=\frac1{16\pi^2}(11-\frac23 N_F)$ and
$\beta_1=\frac1{16\pi^2}(102-\frac{38}3 N_F)$.
Integration of this equation yields the desired relation
\be
 a(g) = \frac{1}{\Lambda_L} (\beta_0 g^2)^{-{\beta_1}/{2\beta_0^2}}\,
e^{-{1}/{2\beta_0g^2}}\label{eq:3.48}
\ee
where $\Lambda_L$ is an integration constant and we have used the values of
$\beta_0$ and $\beta_1$ for $SU(3)$ with $N_F$ Fermions.
 
The constant $\Lambda_L$ governing the relation between the bare coupling
constant and the lattice cutoff can be related by one-loop continuum
calculations to the constants $\Lambda_{\rm MOM}$ and
$\Lambda_{\overline{MS}}$,
which govern the relation between the renormalized coupling constant and
continuum cutoff using the momentum space subtraction procedure in Feynman
gauge and the minimal subtraction procedure respectively, with the
results~\cite{M:JN:5a}
\be
\Lambda_{\rm MOM} = 83.5\,\Lambda_L\ \ ,\qquad \Lambda_{\overline{MS}} =
28.9\,\Lambda_L\ \ .
\label{eq:3.49}
\ee
This correspondence is important for two reasons.  First, the large
coefficients in (\ref{eq:3.49}) allow us to reconcile our notion that the basic scale
$\Lambda_{\rm QCD}$ is of order several hundred MeV with the fact that lattice
measurements yield values of $\Lambda_L\sim 4-4.6$~MeV, which would otherwise
appear astonishingly low.  Second, in principle, it will provide a
quantitative consistency test if experiments in the perturbative regime of QCD
can produce sufficiently accurate values of $\Lambda_{\rm MOM}$ or
$\Lambda_{\overline{MS}}$.
 
There is now convincing evidence that lattice calculations in the
pure gauge sector display the correct renormalization group behavior, and thus
provide accurate solutions of continuum QCD.  Numerical evidence exists for two
independent quantities, $T_{\rm tr}$, the temperature of the deconfinement
transition, and the string tension $\sigma$.   
The
transition temperature on a lattice with $N_t$ time slices is given by $T_{\rm tr} =
{1/ N_t a(g_{\tr})}$ where $g_{\tr}$ is the value of the coupling
for which the transition occurs.  If $a(g_{\rm tr})$ is calculated using the
perturbative expression (\ref{eq:3.48}), then once $g$ is small enough that the
lattice theory coincides with the continuum theory, the quantity $T_{\rm
tr}/\Lambda_L$ should approach a constant.  
A number of calculations show that this is the case
above $\beta = {6/ g^2} = 6$.
Similarly, one observes the same behavior in the string tension
 by plotting
${\sigma/\Lambda^2_L}$ as a function of $\beta = {6/g^2}$ and
observing that  this ratio approaches a constant beyond $\beta=6$.

On the theoretical side, impressive progress has been made in constructing
improved actions, which in addition to containing the leading order contribution
to the action in Eq.~\ref{eq:3.15}, also include corrections to higher order in~$a$ (or
equivalently, $1/g^2$). The most straightforward corrections are calculated
perturbatively, but suffer from the fact that each additional operator in field
theory has its own renormalization constant, and making a 10\% error by
calculating such a constant perturbatively is like making 10\% typing errors in
the coefficients of an allegedly high-order Runge-Kutta integration formula. A
more effective approach is the recent use of renormalization group
methods~\cite{M:JN:1} to derive nonperturbatively corrected ``perfect actions'' which
provide extremely impressive approximations to the continuum theory.

In summary,  lattice gauge theory in the
pure gauge sector is quite satisfactory.  There are no glaring conceptual
or computational problems, and all the numerical evidence to date suggests
that one obtains an excellent approximation to the continuum theory for
$\beta_g$ above 6.  In contrast, we will now see that full QCD including dynamical 
Fermions is more problematic at both the conceptual and computational levels.

\section{Lattice QCD with Quarks}\label{JWN:sec:4}
Significant new problems arise when one attempts to apply the ideas which are
so successful for the stochastic evolution of path integrals for Bosons to
many-Fermion systems.  The fundamental underlying problem is the minus signs
arising from antisymmetry.  In the absence of projection onto the antisymmetric
subspace, $e^{-\beta H}$ filters out the lowest state of any symmetry, and
favors the lowest symmetric state of energy $E^S_0$ relative to the lowest
antisymmetric state with energy $E^A_0$ by the factor $e^{-\beta\left( E^A_0
- E^S_0\right)}$.  If one attempts to project stochastically, for example by
antisymmetrizing path integral Monte Carlo evolution at each step, the
projection error is of order ${1/ \sqrt{N}}$ and can never overcome the
exponential factor in cases in which $E^A_0$ represents a half filled band,
Fermi sea, or Dirac sea and $E^S_0$ corresponds to a Bose condensate in the
lowest state.  Thus, aside from special cases such as one spatial dimension,
 the only known alternative is to write a path integral with Grassmann variables,
introduce integrals over auxiliary fields if necessary to reduce it to
quadratic form, and integrate out the Fermion variables as in Eq.~(\ref{eq:2.26}) to
obtain a Bosonic action containing a Fermionic determinant.  Since the
Grassmann integral has been done analytically, the projection onto the
antisymmetric space is exact, eliminating one part of the sign problem. The
resulting determinant may be positive or negative, so there still remains the
danger of catastrophic sign cancellations in the stochastic evaluation of the
remaining Bosonic integrals.  If, however, as in our present case, there is an
even number of Fermion species with the same action, the determinant appears
with an even power and this final sign problem is also eliminated.
 
This major detour to beat the Fermion sign problem comes at a high price.  We
started with Fermions, which in occupation number representation are
represented by a bunch of 1's and 0's, and we seek to deal with them on a
digital computer which can only work in terms of 1`s and 0's.  Yet we must
resort to calculating determinants of huge, non-local, nearly-singular matrices
involving exceedingly large numbers of floating-point variables and operations.
In addition, by virtue of putting the Fermions  on a lattice, we encounter an
additional unexpected difficulty associated with Fermion doubling, which we
will discuss next.  It should thus be clear at the outset, that the treatment
of Fermions provides fertile ground for new ideas.

\subsection{Naive Lattice Fermions and Doubling}
\begin{figure}
$$\BoxedEPSF{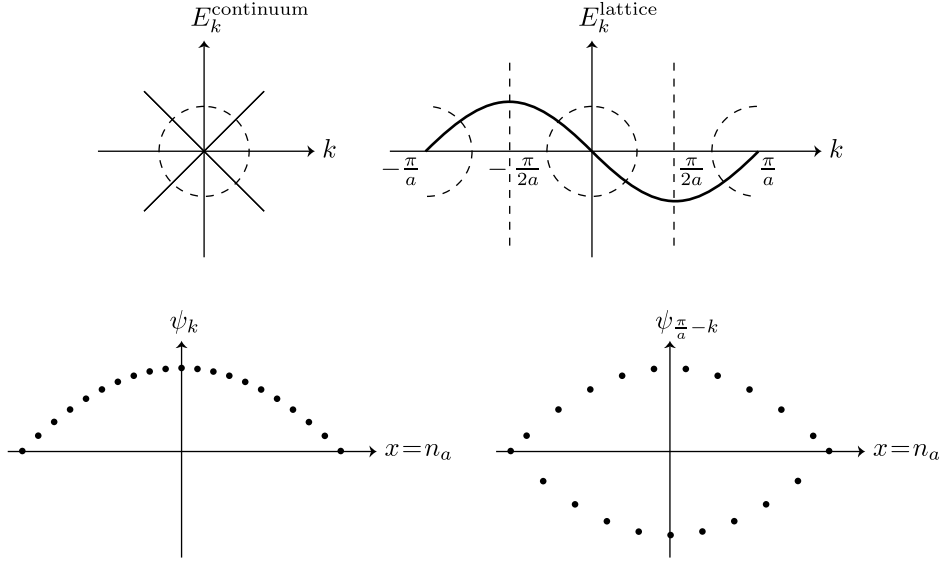}$$
\caption{Fermion doubling in one dimension.
The top plots compare the physical continuum spectrum with the spectrum of the
lattice Hamiltonian.  The lower plots show a half wavelength of the real part
of the non-vanishing component of a physical
wave function $\Psi_k(x)$ and its degenerate unphysical sawtooth partner
$\Psi_{{\pi\over a}-k}(x)$.}
\label{JN:Fig4}
\end{figure}

To appreciate the essential issues, consider the simplest Hermitian finite
difference expression with the desired continuum limit:
\bea
S^{\rm naive}_F &=& a^4 \sum_{\svec n} \biggl[ \bar{\psi} ({\svec
n}) m \psi({\svec n}) + {1\over 2a} \sum_\mu \biggl\{ \bar{\psi} ({\svec n})
\gamma_\mu U_\mu(n) \psi\left( {\svec n} + a_\mu\right) \nonumber\\
&&\qquad{} - \bar{\psi} \left( {\svec n} + a_\mu\right) \gamma_\mu
U^\dagger_\mu (n) \psi ({\svec n}) \biggr\} \biggr] \nonumber\\
&&\Longrightarrow \int d^4x\biggl[ \bar{\psi} m\psi + {1\over 2a} \sum_\mu\biggl\{
\bar{\psi}\gamma_\mu\left( 1 + i gA_\mu\right) \left( 1 +
a\partial_\mu\right)\psi \nonumber\\
&&\qquad{} - \bar{\psi} \left( 1 + a \stackrel{\leftarrow}{\partial\null}_\mu\right)
\gamma_\mu \left( 1 - igaA_\mu\right) \psi\biggr\} \biggr] \nonumber\\
&&\Longrightarrow \int d^4x\,\bar{\psi} \left[ m+\gamma_\mu \left(\partial_\mu + i
g A_\mu\right) \right] \psi \ \ .
\label{eq:4.1}
\eea
Note that throughout we will use Euclidean $\gamma$-matrices satisfying
$\gamma_\mu \gamma_\nu + \gamma_\nu\gamma_\mu = 2\delta_{\mu\nu}$, for which
an explicit representation is given in Ref.\cite{R:JN:06}.  Although this naive action
appears to have the desired continuum behavior and symmetries, it has an
unexpected problem.  To see this problem in its simplest form, consider the
Hamiltonian corresponding to the action (\ref{eq:4.1}) in one space dimension for
the special case of free quarks ($A=0$) and zero mass:
\bea
H^{\rm naive} &=& a\sum_n \psi^\dagger(n)
\alpha {1\over i2a} \left(
\psi{(n+1)} - \psi{(n-1)}\right) \nonumber\\
&&\Longrightarrow \int dx\,\psi^\dagger\alpha {1\over i} \partial_x \psi 
\label{eq:4.2}
\eea
where $\alpha = \gamma_0\gamma_1$ and for convenience we choose the
representation $\alpha = \left(\matrix{ 1 & 0 \cr 0 & -1\cr}\right)$.  Note
that $H^{\rm naive}$ is Hermitian by virtue of the symmetric difference
$\psi{(n+1)} - \psi{(n-1)}$.  We now transform the field operators
to momentum space by writing the
Fourier sum
\be
\psi(n) = {1\over \sqrt{Na}} \sum\limits^{\pi\over a}_{k = -{\pi\over a}}
\psi_k e^{ik\,na} \label{eq:4.3a}
\ee
where it is understood that for a lattice with $N$ sites and periodic boundary
conditions, the sum over momenta in the first Brillouin zone extends over the
$N$ momenta
\be
k_p = {p\pi\over Na}\qquad - {N\over 2} \le p \le {N\over 2}\ \ . 
\label{eq:4.3b}
\ee
The Hamiltonian is diagonal
\be
H^{\rm naive} = \sum\limits^{\pi\over a}_{k = -{\pi\over a}} \psi^\dagger_k
\alpha{\sin(ka)\over a} \psi_k \label{eq:4.4a}
\ee
and thus has the eigenvalue spectrum
\be 
E_k = \pm {\sin ka\over a}\mathop{\sim}\limits_{k\to 0} \pm k \Bigl( 1 -
{(ka)^2\over 3} +\cdots\Bigr)  \label{eq:4.4b}
\ee
with eigenfunctions
\be
\Psi^\pm_k(n) = e^{ikna} \chi^\pm
\ee
where $\chi^\pm$ is a two component spinor with either an upper or lower
component unity and the other component zero.
The comparison of the continuum spectrum for a massless Dirac particle
$E_k=\pm k$ and the lattice spectrum in the top of Fig.~\ref{JN:Fig4} displays the
species doubling problem.  In the region denoted by the dashed circle centered
at the origin, the lattice spectrum (\ref{eq:4.4b}) yields a good approximation to the
linear physical spectrum, and the range of linearity increases as $a\to 0$.
However, at the edge of the Brillouin zone, there is a second region in which
the spectrum also goes to zero linearly, denoted by the two dashed
semicircles.  In fact, for every physical mode $\Psi_k$, there is a precisely
degenerate unphysical mode $\Psi_{{\pi\over a}-k}$.  Since the partition
function blindly counts and weights all modes according to their energies, it
is clear that all Fermion loops will be overcounted by a factor of 2 in all
physical observables.  Note also that since the velocity is $v = {dE\over
dk}$, the lattice spectrum necessarily mixes right-moving and left-moving
modes.

The origin and structure of the doubled states is simple. The degenerate
partner to the state $\Psi_k(n) = e^{ik\,na}\chi$
is the state $\psi_{{\pi\over a} -
k}(n)= e^{in\pi} e^{- ik\,na}\chi$,
that is, a sawtooth mode in which every other
lattice site has an extra factor of $-1$.  The real part of a low $k$ mode
$\Psi_k$ and its sawtooth partner $\Psi_{{\pi\over a}-k}$ are sketched for a
half wavelength in the lower section of Fig.~\ref{JN:Fig4}.  Note that although there
are sufficient points in the half wavelength of $\Psi_k$ to yield an accurate
integral with any smooth function, there is no way that the rapidly
oscillating wave function $\Psi_{{\pi\over a} - k}$ can represent a mode with
momentum near ${\pi\over a}$.  Thus, we need some way to eliminate these modes
so that they play no role in the continuum limit.  The origin of the
degeneracy of the physical mode with its sawtooth partner is the symmetric
difference approximation to the derivative $\psi'\sim {\psi{(n+1)} -
\psi{(n-1)}\over 2a}$ in the naive Hamiltonian (4.2), which clearly is
impervious to the minus signs $e^{- i n \pi}$ and thus yields the same
magnitude for the derivatives of $\psi_k$ and $\psi_{{\pi\over a}-k}$.  The
origin of this symmetric difference, in turn, is Hermiticity, since
expressions involving only nearest neighbor differences like
$\psi^\dagger n(\psi{(n+1)} - \psi(n))$ are non-Hermitian and yield complex
eigenvalues.
 
We may note in passing that a finite difference approximation to a second
derivative strongly breaks the degeneracy between the physical and sawtooth
modes.  Indeed, a perturbation of the following form
\bea
H' &=& - \sum\limits_n \psi^\dagger(n)\gamma_0 \left( \psi{(n+1)} - 2\psi(n) +
\psi{(n-1)}\right) \nonumber\\
&=& - a\,a\sum\limits_n \psi^\dagger(n)\gamma_0\Bigl( {\psi{(n+1)} - 2\psi(n) +
\psi{(n-1)}\over a^2}\Bigr) \nonumber\\
&&\underarrow{a\to 0} - a \int dx\,\bar{\psi}(x)
\psi''(x) \underarrow{a\to 0}0 \label{eq:4.5a}
\eea
which vanishes in the continuum limit and behaves like a momentum-dependent
mass term
\be
H' = \sum\limits_k \bar{\psi}_k m (k) \psi_k \label{eq:4.5b}
\ee
where
\bea
m(k) &=& {2\over a}\left( 1 - cos (ka)\right) \nonumber\\
&&\mathop{\sim}\limits_{k\to 0} ak^2\nonumber\\
&&\mathop{\sim}\limits_{k\to{\pi\over a}} {4\over a}\ \ .
\label{eq:4.5c}
\eea
For the physical modes in the region of the origin, the perturbation has no
effect as $a\to 0$, consistent with the fact $H'$ vanishes in the continuous
limit.  For the spurious sawtooth modes, however, the mass diverges as
${1\over a}$.  Thus, one way to remove the unphysical modes would be to include
a term of the form (\ref{eq:4.5a}) in the lattice Hamiltonian to raise their mass so
high that they contribute negligibly to the partition function, and this is
the idea underlying Wilson's lattice action for Fermions discussed below.
 
The doubling we have discussed for simplicity in one dimension arises
analogously in each of the
four Euclidean dimensions of the naive Fermion action, Eq.~(\ref{eq:4.1}), so that we
obtain $2^4=16$ lattice modes for each physical mode.   Again, specializing to
the massless case, the momentum space action corresponding to Eq.~(\ref{eq:4.1}) is
\be
S^{\rm naive}_F \equiv \bar{\psi}M\psi = \sum_k \bar{\psi}_k \sum_\mu
\gamma_\mu {\sin\left( k^\mu a\right)\over a} \psi_k \label{eq:4.6a}
\ee
so that the inverse propagator is
\be
\ll T\bar{\psi}\psi\rr^{-1} = M(k) = \sum_\mu \gamma_\mu {\sin\left( k^\mu
a\right)\over a} \ \ .
\label{eq:4.6b}
\ee
This propagator
replicates the physical behavior in the region of $k^\mu\sim 0$ fifteen times
around points on the edge of the four-dimensional Brillouin zone at which one
or more of the components $k^\mu\sim {\pi\over a}$.

We are now prepared to understand both the features giving rise to
the doubling problem and the generality of the problem.\cite{M:JN:13}  Whereas the
specific function $\sin\left(k^\mu a\right)$ in Eqs.~(4.4a) and (4.6) is the
result of using the lowest-order Hermitian difference formula for the
derivatives in the continuum action, the most general form of the chiral
symmetric, Hermitian action derived from discrete derivatives on a periodic
lattice with the correct continuum limit is
\be
S_F = \sum_k \bar{\psi}_k \sum_\mu \gamma_\mu P^\mu(k) \psi_k 
\label{eq:4.7}
\ee
where $P^\mu(k)$ is real for Hermiticity, $P^\mu(k)\underarrow{k\to 0} 0$ for
the correct continuum limit, and $P^\mu(k)$ is periodic under $k^\mu\to k^\mu
+ {2\pi\over a}$ and continuous for local discrete difference formulae on a
lattice.  Note that chiral symmetry requires the form $\bar{\psi} \gamma_\mu
P^\mu\psi$, so that under a chiral transformation $\psi\to
e^{i\alpha\gamma_5}\psi$, the two sign changes from $\gamma_0$ in
$\bar{\psi}$ and $\gamma_\mu$ leave the action invariant.  Since $P^\mu(k)$ is
real, continuous and periodic in $k^\mu$ with period ${2\pi\over a}$, it must
cross the axis at some intermediate point, so that this general discrete
action has the doubling observed in Eq.~(\ref{eq:4.6a}) in each of the four Euclidean
directions yielding 15 spurious low-mass excitations for each physical
excitation.  A rigorous version of these arguments is known as the
Nielsen--Ninomiya no-go theorem,\cite{M:JN:14} which proves using homotopy
theory that one cannot avoid Fermion doubling in a lattice theory which is
simultaneously Hermitian, local and chiral symmetric.
 
An additional aspect of Fermion doubling is the absence of the axial anomaly.
The axial current, which is conserved for gauge theories at the classical
level but not conserved at the quantum level in the continuum theory, is
conserved for naive lattice Fermions.  Again, the culprits are the unphysical
lattice duplicates of the physical Fermion excitations, which couple to an
external axial current with the opposite chiral charge and effectively cancel
the axial anomaly arising from the physical Fermions.

\subsection{Wilson Fermions}
One of the ways out of the no-go theorem is to give up chiral symmetry and,
following Wilson, add a second derivative term of the form (\ref{eq:4.5a}) to raise
the mass of the unphysical sawtooth modes.  In one dimension, combining the naive
Hamiltonian (\ref{eq:4.2}) with a multiple $r$ of the perturbation (\ref{eq:4.5a}) yields
the Wilson Hamiltonian
\bea
H_{\rm W} &=& a \sum_n\psi^\dagger(n) \Bigl[ {\alpha\over i}
{\left(\psi(n+1) - \psi(n-1)\right) \over 2a} - {ra\gamma_0\over 2i}\
{\psi(n+1) - 2\psi(n) + \psi(n-1)\over a^2} \Bigr] \nonumber\\
&&=\sum_k \psi^\dagger_k \Bigl[ \alpha {\sin(ka)\over a} - r
{\gamma_0\over i} {\left(\cos(ka) - 1\right)\over a}\Bigr] \psi_k\ \ .
\label{eq:4.8a}
\eea
Using a representation with $\alpha = \sigma_3$ and $ - i \gamma_0 =
\sigma_1$, we obtain the energy spectrum
\be
E^2 = \Bigl( {\sin (ka)\over a}\Bigr)^2 + \Bigl[ {r\over a} \Bigl( \cos
(ka) - 1 \Bigr)\Bigr]^2 \label{eq:4.8b}
\ee
with the limits
\bea
E&&\underarrow{k\to 0} \pm k \Bigl( 1 - \fracs16 k^2 a^2 +
{r^2\over 8} k^2 a^2\Bigr) \nonumber\\
E&&\underarrow{k\to {\pi\over a}} \pm {2r\over a}\ \ .
\label{eq:4.8c}
\eea
Thus, for fixed $r$, the mode for $k\sim 0$ has the correct continuum limit
whereas the sawtooth mode for $k\sim {\pi\over a}$ becomes infinitely massive
and decouples from the theory.
 
In four Euclidean dimensions, the corresponding Wilson action is
\bea
 S_{\rm W} &=& -a^4 \sum_{\svec n} {1\over 2a} \sum_\mu\left[
\bar{\psi} ({\svec n}) (r-\gamma_\mu) U_\mu ({\svec n}) \psi ({\svec n}+a_\mu)
+ \bar{\psi} ({\svec n}+a_\mu) (r+\gamma_\mu) U^\dagger_\mu({\svec n})
\psi({\svec n})\right]\nonumber\\
&&\quad{}+ a^4 \sum_{\svec n}\Bigl( m+{4r\over a}\Bigr)\bar{\psi}({\svec n})
\psi({\svec n}) 
\label{eq:4.9}
\eea
and the propagators for the spurious modes acquire masses which diverge as
${r\over a}$ as in the one-dimensional case.
 
The Wilson action manifestly breaks chiral symmetry for $m=0$, since under the
transformation $\psi\to e^{i\alpha\gamma_5}\psi$, $\bar{\psi}\psi \to
\bar{\psi}\,e^{i2\alpha\gamma_5}\psi$.  As long as the contribution of the
symmetry breaking term can be made arbitrarily small, its presence does not
interfere with the physics of spontaneous symmetry breaking.  In the case of a
spin system, for example, one defines the spontaneous magnetization as the
limit of the thermodynamic trace of the spin in the presence of an external
magnetic field in the limit as the external field goes to zero.
Heuristically, we expect the contribution of the chiral symmetry breaking term
to vanish in the continuum limit even though the masses of the unphysical
modes go to infinity since $a\int dx\,\bar{\psi}\Sq^2 \psi
\underarrow{a\to 0} 0$, and explicit calculation~\cite{M:JN:13} shows that this is the
case.  Furthermore, including 15 massive Wilson Fermion partners as well as the
physical mode yields the correct axial anomaly in the continuum limit.\cite{M:JN:13}
 
The combination $M=\left( m+{4r\over a}\right)$ in Eq.~(\ref{eq:4.9}) enters the
action like a mass term.  Since this mass term is not protected from renormalization
by any symmetry, it must be renormalized by ``fine tuning.''  Thus, by a
search involving a series of calculations of the pion mass at fixed coupling
constant $g$ and Wilson $r$, we may determine a critical mass $M_{\rm
cr}(g,r)$ such that $m_\pi = 0$ for a chiral symmetric theory
and $M(g,r)$ such that $m_\pi = 140$~MeV for the physical theory.  To
the extent to which it is meaningful to define a quark mass, one may define
$M_{\rm quark}\equiv M - M_{\rm cr}$.  A more conventional notation is in
terms of the hopping parameter
\be
\kappa\equiv {1\over 2Ma} = {1\over 2ma + 8r} 
\label{eq:4.10a}
\ee
and rescaled Fermion fields
\be
\Psi \equiv \left( Ma^4\right)^{1/2} \psi
\label{eq:4.10b}
\ee
for which the action has the form
\bea
S_{\rm W} &=& \sum_{\svec n} \biggl\{ \bar{\Psi} ({\svec n})
\Psi({\svec n}) - \kappa\sum_\mu \Bigl[ \bar{\Psi} ({\svec n}) (r-\gamma_\mu)
U_\mu ({\svec n}) \Psi ({\svec n}+a_\mu) \nonumber\\
&&\qquad {} + \bar{\Psi} ({\svec n}+a_\mu) (r+\gamma_\mu)U^\dagger_\mu({\svec
n})
\Psi({\svec n}) \Bigr] \biggr\} 
\label{eq:4.10c}
\eea
and
\be
M_{\rm quark} = {1\over 2a} \Bigl( {1\over \kappa}
- {1\over \kappa_{\rm cr}}\Bigr)\ \ .
\label{eq:4.10d}
\ee
The fields have been scaled such that the diagonal term is now unity, and the
hopping parameter $\kappa$
specifies the strength of the nearest-neighbor coupling
via link variables.  Although the Wilson parameter $r$ is often chosen to be
1, in principle it may be optimized to render the errors from
$ar\bar{\psi}\Sq^2\psi$ in the physical modes and the contribution of the
sawtooth modes comparable.
 
Just as one can derive the conserved vector current of the continuum theory as
the Noether current of the continuum action, one can derive a discrete version
of Noether's theorem from the Wilson action (\ref{eq:4.9}) of the form
\be
\Delta^\mu V_\mu({\svec n}) \equiv V_\mu({\svec n}) - V_\mu \left({\svec
n}-a_\mu\right) = 0 
\label{eq:4.11a}
\ee
where the conserved vector current on the lattice is
\bea
V_\mu({\svec n}) &=&- \fracs12\bar{\psi} ({\svec n})
(r-\gamma_\mu) U_\mu ({\svec n})\psi({\svec n}+a_\mu) \nonumber\\
&&{}+\fracs12 \bar{\psi} ({\svec n}+\mu) (r+\gamma_\mu) U^\dagger_\mu({\svec
n}) \psi({\svec n}) \ \ . 
\label{eq:4.11b}
\eea
Note that neither the local current $\bar{\psi}({\svec n})
\gamma_\mu\psi({\svec n})$ nor the point-split current defined by
(\ref{eq:4.11b}) with
$r=0$ is conserved on the lattice.  In contrast, since there is no chiral
symmetry, there is no unique definition of the lattice axial current, so it is
necessary to use perturbation theory to explicitly calculate the difference
between any choice for the lattice axial current and the continuum axial
current.

In summary, Wilson Fermions provide one possible framework for solving the
doubling problem on the lattice and calculations discussed subsequently will be based on
this formulation. Other alternatives include staggered fermions~\cite{M:JN:15}, which
while not avoiding the no-go theorem, maintain a remnant of chiral symmetry while
thinning the spurious degrees of freedom from 16 down to 4 in four dimensions,
and Kaplan--Shamir fermions~\cite{M:JN:16}, which are formulated on the
four-dimensional boundary of a five-dimensional sphere.

\subsection{Hopping parameter expansion}

Just as the strong-coupling expansion provided insight into the solutions of lattice QCD in
the pure gauge sector, expansion in powers of the hopping parameter, $K$, provides
analogous insights into solutions in the presence of Fermions. The basic idea is to expand
the integral
\begin{equation}
Z = \int {\cal D}(\bar\psi \psi) {\cal D}(U) e^{-\bar\psi(1+KU)\psi - S(U)}
\end{equation}
in powers of the hopping term $\bar\psi KU\psi_{n\pm1}$. Integration over all
$\bar\psi\psi$ then yields the sum of all possible closed chains of $KU$ in which the
$U$'s are oriented head to tail, and these closed quark paths represent all the quark time
histories. As before, integrating over all $U$ then tiles the closed quark loops with gluons,
giving rise to all the color singlet flux tubes connecting the quarks and antiquarks on any
time slice.

Now, consider a path integral corresponding to the propagation of a meson from
space-time point~$x$ to~$y$, which we write schematically, ignoring $\gamma$ matrices,
as 
\begin{eqnarray}
\ll T e^{-\beta H}\bar\psi_y \psi_y \bar\psi_x \psi_x\rr
&=& Z^{-1}\int{\cal D}(U){\cal D}(\bar\psi\psi) e^{-\bar\psi M(U)\psi - S(U)}
\bar\psi_y \psi_y \bar\psi_x \psi_x \nonumber\\
&=& Z^{-1}\int{\cal D}(U) e^{\ln\det M(U)-S(U)} M^{-1}_{yx}(U) M^{-1}_{xy}(U) 
\end{eqnarray}
where $M=(1+KU)$. Expanding $M^{-1}_{yx}(U) = (1+KU)^{-1}_{yx}  $ in powers of $K$
generates all valence quark trajectories from $x$ to $y$ and similarly $M^{-1}_{xy}(U)$
generates trajectories from $y$ to $x$. 

\begin{figure}
$$\BoxedEPSF{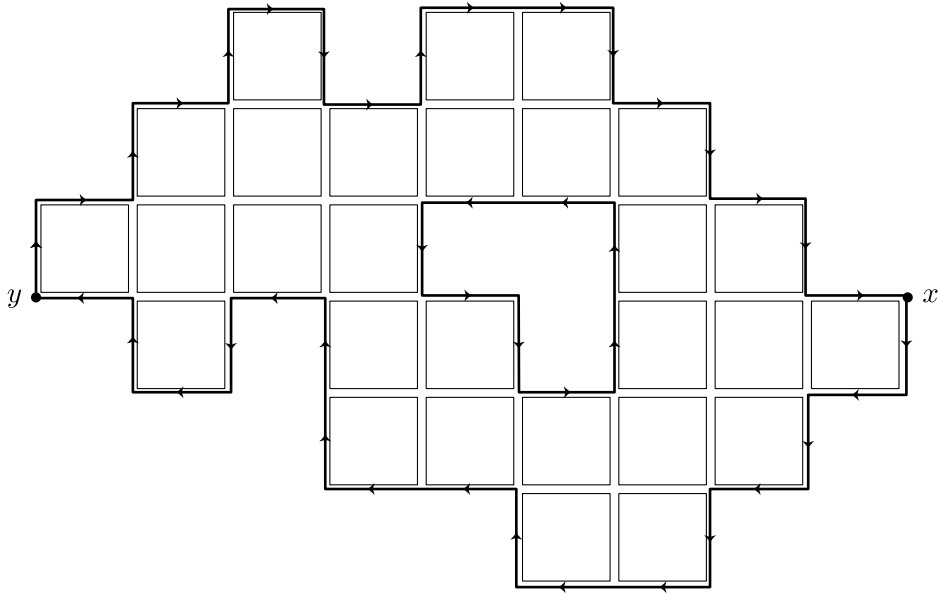}$$
\caption{A typical time history arising in the hopping parameter expansion. The
heavy lines connecting $x$ and $y$ denote valence quarks, the interior heavy line
corresponds to a quark-antiquark pair, and the light lines denote gluons.}
\label{JN:Fig5}
\end{figure}

Expansion of $\ln\det  M(U)$ generates all disconnected quark loops corresponding to
excitation of quark-antiquark pairs from the Dirac sea, and expansion of $S(U)$ tiles
surfaces between these valence and sea-quark trajectories. A typical configuration
showing a valence quark-antiquark pair propagating from $x$ to $y$,
a quark-antiquark loop excited out of the fermi sea, and a minimal tiling by
plaquettes is shown in Fig.~5. Cutting this tiling at various time slices corresponds to
a quark-antiquark pair connected by a flux tube or two quark-antiquark pairs
connected by flux tubes. From this argument, we see that omission of the
determinant, which is very expensive computationally, yields the quenched or
valence approximation in which quark-antiquark pairs excited from the sea are
neglected. Typical lattices used in numerical calculation vary from $16^2\times 32$
to $32^3\times64$ sites and thus involve integration over $10^7$ to $10^8$ real
variables.


\subsection{Correlation functions}
As in the case of other strongly interacting many-body systems, to
understand the structure of the vacuum and light hadrons in
nonperturbative QCD, it is instructive to study appropriately selected
ground state correlation functions, to calculate their properties
quantitatively, and to understand their behavior physically.

Because of our subsequent interest in instantons, we will focus our attention on
vacuum  point-to-point equal time
correlation functions of hadronic currents
\begin{equation}
R(x) = \langle \Omega |T J(x) \bar{J} (0)|\Omega \rangle
\label{E:JN:3}
\end{equation}
discussed in detail by Shuryak~\cite{R:JN:03} and recently calculated in
quenched lattice QCD.\cite{R:JN:07} The motivation for supplementing
knowledge of hadron bound state properties by these correlation
functions is clear if one considers the deuteron. Simply knowing the
binding energy, rms radius, quadruple moment and other ground state
properties yields very little information about the nucleon-nucleon
interaction in each spin, isospin and angular momentum channel as a
function of spatial separation.  To understand the nuclear interaction in
detail, one inevitably would be led to study nucleon-nucleon scattering
phase shifts.  Although, regrettably, our experimental colleagues have
been most inept in providing us with quark-antiquark phase shifts, the
same physical information is contained in the vacuum hadron current
correlation functions $R(x)$.  As shown by Shuryak~\cite{R:JN:03}, in many
channels these correlators may be determined or significantly
constrained from experimental data using dispersion relations.  Since
numerical calculations on the lattice agree with empirical results where
available, we regard the lattice results as valid solutions of QCD in all
channels and thus use them to obtain information comparable to
scattering phase shifts.

\begin{figure}[htb] 
\begin{center}
\BoxedEPSF{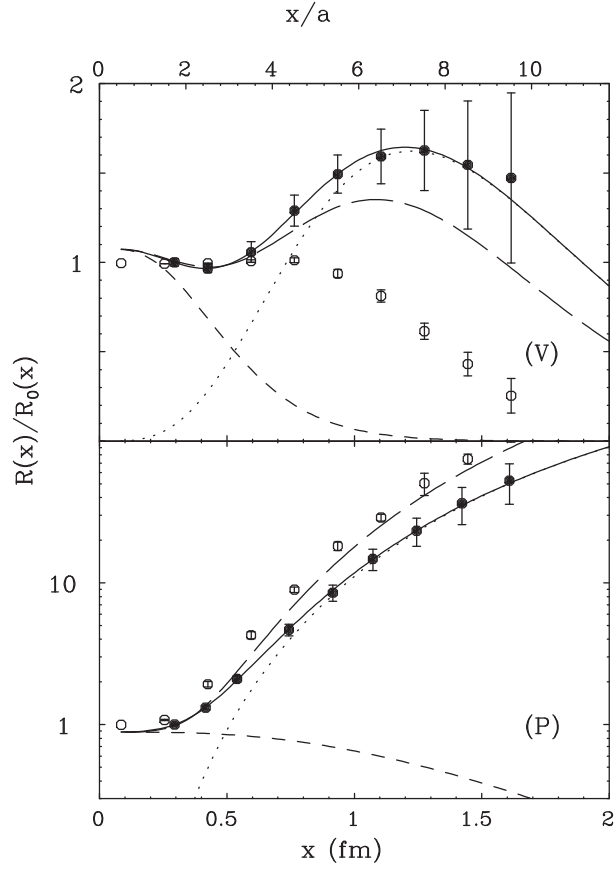}
\medskip
\caption{Vector (V) and 
Pseudoscalar (P) correlation functions are shown in the upper and lower panels
respectively.  Lattice results~\cite{R:JN:07} are denoted by the solid points with
error bars and fit by the solid curves, which may be decomposed into continuum and
resonance components denoted by short dashed and dotted curves respectively. 
Phenomenological results determined by dispersion analysis of
experimental data in Ref.~\cite{R:JN:03} are shown by long dashed
curves, and the open circles denote the results of the random instanton
model of Ref.~\cite{R:JN:04}.}
\label{F:JN:1}  
\end{center}
\end{figure}

The correlation functions we calculate in the pseudoscalar, vector,
nucleon and Delta channels are
\begin{eqnarray*}
R(x) &=& \langle \Omega |T J^p(x) \bar{J}^p (0)|\Omega \rangle \\
R(x) &=& \langle \Omega |T J_\mu(x) \bar{J}_\mu (0)|\Omega \rangle 
\\ R(x) &=& {\textstyle\frac14} \Tr \left(\langle \Omega |T J^N(x) \bar{J}^N
(0)|\Omega \rangle x_\nu \gamma_\nu \right) \\
\noalign{\hbox{and}}
R(x) &=& {\textstyle\frac14} \Tr \left(\langle \Omega |T J^\Delta_\mu (x)
\bar{J}^\Delta_\mu (0)|\Omega \rangle x_\nu \gamma_\nu  \right)
\end{eqnarray*}
where
\begin{eqnarray*}
J^p &=& \bar{u} \gamma_5 d  \\
J_\mu &=& \bar{u} \gamma_\mu \gamma_5 d   \\
J^N &=& \epsilon_{abc} [u^a C \gamma_\mu u^b] \gamma_\mu \gamma_5
d^c 
\\
\noalign{\hbox{and}}
J^\Delta_\mu &=& \epsilon_{abc} [u^a C \gamma_\mu u^b] u^c \  . 
\end{eqnarray*}
As in Refs.~\cite{R:JN:03} and~\cite{R:JN:07}, we consider the ratio of the
correlation function in QCD to the correlation function for non-interacting
massless quarks,
${R(x)}/{R_0 (x)}$, which approaches one as $x \to 0$ and displays a
broad range of non-perturbative effects for $x$ of the order of 1~fm. 
Typical results of lattice calculations of ratios of vacuum correlation
functions are shown in Fig.~\ref{F:JN:1}.

Note that the lattice results (solid line) agree well with phenomenological results
from dispersion analysis of data (long dashed curves).  Also, observe that the
vector and pseudoscalar correlation functions are strongly dominated by the rho
and pion contributions (dotted lines) in the region of 0.5 to 1.5 fm.  We will
subsequently show that these rho and pion contributions in turn arise from the
zero mode contributions associated with instantons.

As discussed in Refs.~\cite{R:JN:03,R:JN:07}, these vacuum correlators show strong
indications of instanton dominated physics.  As shown by 't~Hooft~\cite{R:JN:08},
the instanton induced interaction couples quarks and antiquarks of opposite
chirality leading to strong attractive and repulsive forces in the pseudoscalar and
scalar channels respectively and no interaction to leading order in the vector
channel.  Just this qualitative behavior is observed at short distance in all the
channels we computed.  Furthermore, as shown by the open circles with error
bars   in Fig.~\ref{F:JN:1}, the random instanton model of Shuryak et
al.\null~\cite{R:JN:04} reproduces the main features of the correlation functions at
large distance as well.

\section{The Role of Instantons in Light Hadrons}\label{JWN:sec:5}
Having established the framework of lattice QCD, I will now use it as a tool to elucidate
the role of instantons in light hadrons.
The QCD vacuum is understood as a superposition of
an infinite number of states of different winding number, where the winding number
characterizes the number of times the group manifold is covered when one covers the
physical space. Just as there is a stationary point in the action of the Euclidean Feynman
path integral for a double well potential corresponding to the tunneling between the two
degenerate minima, so also there is a classical solution to the QCD equations in
Euclidean time, known as an instanton~\cite{R:JN:09}, which describes tunneling
between two vacuum states of differing winding number.  The action associated
with an instanton is
\begin{equation}
S_0 = \fracs14 \int d^4 x\, F^a_{\mu \nu} F^a_{\mu \nu} = \frac{48}{g^2
\rho^4} \int d^4 x\, \Bigl(\frac{\rho^2}{x^2 +\rho^2} \Bigr)^4 = \frac{8 \pi^2}{g^2}\  .
\label{E:JN:4}
\end{equation}
Note that the action density has a universal shape characterized by a size $\rho$, and
that the action is independent of $\rho$.  Furthermore, the instanton field strength is
self-dual, {\it i.e.\/} $\tilde{F}^a_{\mu \nu} \equiv \epsilon_{\mu \nu \alpha \beta}
F^a_{\alpha \beta} = \pm F^a_{\mu \nu}$, so that the topological change of an
instanton is
\begin{equation}
Q \equiv \frac{g^2}{8 \pi^2} {\textstyle\frac14} \int d^4x \tilde{F}^a_{\mu \nu} F^a_{\mu \nu}
= \pm 1\  .
\end{equation}
Two features of instantons are particularly relevant to light hadron physics. The
first is the fact that although the fermion spectrum is identical at each minimum
of the vacuum, quarks of opposite chirality are raised or lowered one level
between adjacent minima.   Thus, an instanton absorbs a left-handed quark of
each flavor and emits a right-handed quark of each flavor, and an anti-instanton
absorbs right-handed quarks and emits left-handed quarks.  Omitting heavier
quarks for simplicity, the resulting 't~Hooft interaction involving the operator
$\bar{u}_R u_L
\bar{d}_R d_L \bar{s}_R s_L$ is the natural mechanism to describe otherwise
puzzling aspects of light hadrons.  It is the natural mechanism to flip the helicity of
a valence quark and transmit this helicity to the glue and quark-antiquark pairs,
thereby explaining the so-called ``spin crisis."  It also explains why the two
valence
$u$ quarks in the proton would induce twice as many  $\bar{d} d$ pairs as the
$\bar{u} u$ pairs induced by the single valence $d$ quark.  The second feature is
that each instanton gives rise to a localized zero mode of the Dirac operator
$D_\mu
\gamma_\mu \phi_0 (x) = 0$.  Hence, considering a spectral  representation of the
quark propagator, it is natural that the propagator for the light quarks is
dominated by these zero modes at low energy.  This gives rise to a physical picture
in which
$\bar{q} q$ pairs propagate by ``hopping"  between localized modes
associated with instantons.

\subsection{Identifying instantons by cooling}
The Feynman path integral for a quantum mechanical problem with degenerate
minima is dominated by paths that fluctuate around stationary solutions to the
classical Euclidean action connecting these minima.\cite{R:JN:10} In the case of the
double well potential, a typical Feynman path is composed of segments fluctuating
around the left and right minima joined by segments crossing the barrier.  If one
had such a trajectory as an initial condition, one could find the nearest stationary
solution to the classical action numerically by using an iterative local relaxation
algorithm.  In this method, which has come to be known as cooling, one
sequentially minimizes the action locally as a function of the coordinate on each
time slice and iteratively approaches a stationary solution.  In the case of the
double well, the trajectory approaches straight lines in the two minima joined by
kinks and anti-kinks crossing the barrier and the structure of the trajectory can be
characterized by the number and positions of the kinks and anti-kinks.

In QCD, the corresponding classical stationary solutions to the Euclidean action for the
gauge field connecting degenerate minima of the vacuum are instantons, and we
apply the analogous cooling technique~\cite{R:JN:11} to identify the instantons
corresponding to each gauge field configuration. 

\begin{figure}[htb] 
\begin{center}
\BoxedEPSF{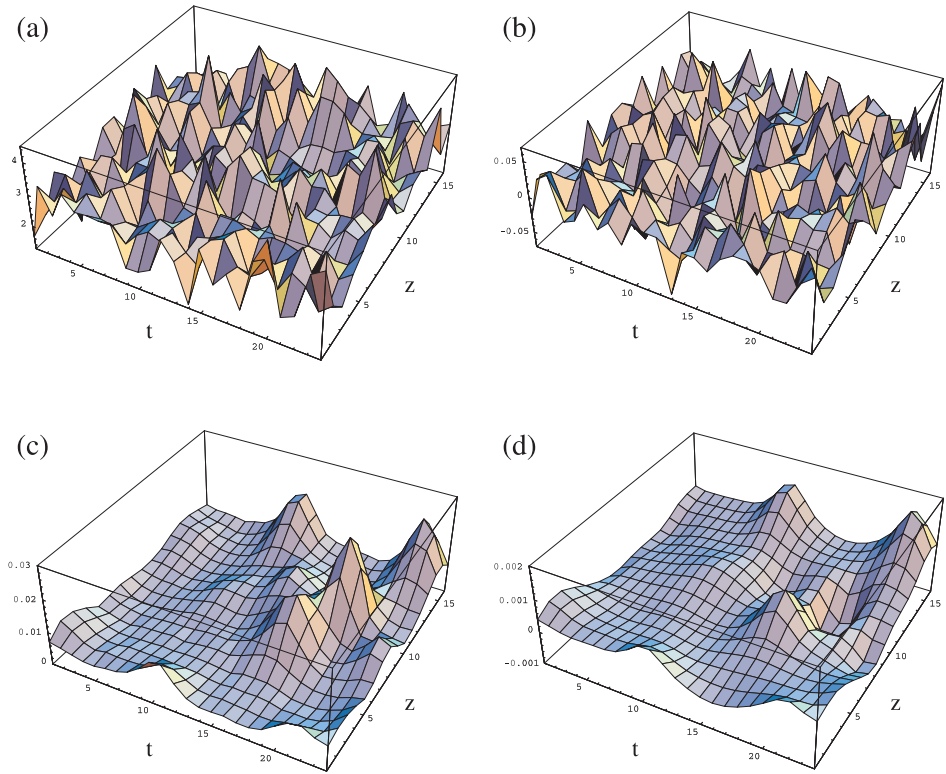}
\medskip
\caption{Instanton content of a typical slice of a gluon configuration at fixed $x$
and $y$ as a function of $z$ and $t$.  The left column shows the action density
$S(1,1,z,t)$ before cooling (a) and after cooling for 25 steps (c).  The right column
shows the topological charge density $Q (1,1,z,t)$ before cooling (b) and after cooling
for 25 steps.}
\label{F:JN:2}  
\end{center}
\end{figure}

The results of using 25 cooling steps as a filter to extract the instanton content of a
typical gluon configuration are shown in Fig.~\ref{F:JN:2}, taken from
Ref.~\cite{R:JN:12} using the Wilson action on a $16^3 \times 24$ lattice at
${6}/{g^2} = 5.7$.  As one can see, there is no recognizable structure before
cooling.  Large, short wavelength fluctuations of the order of the lattice spacing
dominate both the action and topological charge density.  After 25 cooling steps,
three instantons and two anti-instantons can be identified clearly.  The action
density peaks are completely correlated in position and shape with the topological
charge density peaks for instantons and with the topological charge density valleys
for anti-instantons.  Note that both the action and topological charge densities are
reduced by more than two orders of magnitude  so that the fluctuations removed by
cooling are several orders of magnitude larger than the topological excitations that
are retained.

Setting the coupling constant, or equivalently, the lattice spacing, and quark mass by
the nucleon and pion masses in the usual way, it turns out that the characteristic size
of the instantons identified by cooling is 0.36~fm and the density is 1.6 fm$^{-4}$, in
reasonable agreement with the value of 0.33~fm and 1.0~fm$^{-4}$ in the liquid
instanton model.\cite{R:JN:04}

\subsection{Comparison of results with all gluons and with only instantons}\noindent
One dramatic indication of the role of instantons in light hadrons is to
compare observables calculated using all gluon contributions with those obtained
using only   the instantons remaining after cooling.  Note that there are truly dramatic
differences in the gluon content before and after cooling.  Not only has the action
density decreased by two orders of magnitude, but also the string tension has
decreased to 27\% of its original value and the Coulombic and magnetic hyperfine
components of the quark-quark potential are essentially zero.  Hence, for example,
the energies and wave functions of charmed and $B$ mesons would be drastically
changed.

\begin{figure}[htb] 
\begin{center}
\BoxedEPSF{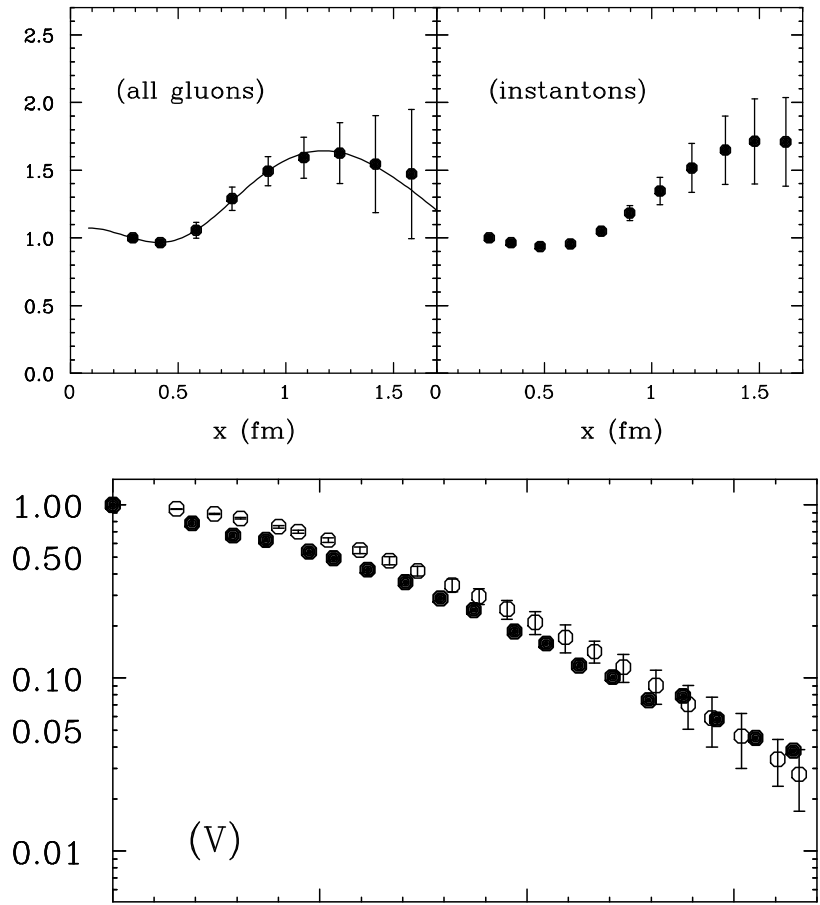}
\caption{Comparison of rho observables calculated with all gluon
configurations and only instantons. The upper left-hand plot shows the vacuum
correlator in the rho channel calculated with all gluons as in
Fig.~\protect\ref{F:JN:1} and the upper right-hand plot shows the analogous result
with only instantons.  The lower plot shows the ground state density-density
correlation function for the rho with all gluons (solid circles) and with only
instantons (open circles).  Error bars for the solid circles are comparable to the
open circles and have been suppressed for clarity.}
\label{F:JN:3}  
\end{center}
\end{figure}

As shown in Fig.~\ref{F:JN:3}, however, the properties of the rho meson are virtually
unchanged.  The vacuum correlation function in the rho (vector) channel and the
spatial distribution of the quarks in the rho ground state, given by the ground state
density-density correlation function~\cite{R:JN:13} $\langle \rho| \bar{q} \gamma_0
q(x) \bar{q} \gamma_0 q(0) | \rho \rangle$, are statistically indistinguishable before
and after cooling.  Also, as shown in Ref.~\cite{R:JN:12}, the rho mass is unchanged
within its 10\% statistical error.  In addition, the pseudoscalar, nucleon, and
delta vacuum correlation functions and nucleon and pion density-density correlation
functions are also qualitatively unchanged after cooling, except for the removal of the
small Coulomb induced cusp at the origin of the pion.

Although these cooling studies strongly indicate that instantons play an essential role
in light quark physics, cooling has the disadvantage of modifying the instanton
content of the original gluon configuration. It is possible to avoid the gradual
shrinkage of a single instanton until it eventually falls through the lattice by using
an improved action that is sufficiently scale independent.\cite{R:JN:14}  However,
pairs of instantons and anti-instantons will eventually attract each other and
annihilate, thereby continually eroding the original distribution.  Hence,  it is
valuable to complement these cooling calculations by studies of the zero modes
associated with instantons, which, as we show in the next section, can be carried out
successfully on the original uncooled gluon configurations.

\begin{figure}[h!] 
\begin{center}
\BoxedEPSF{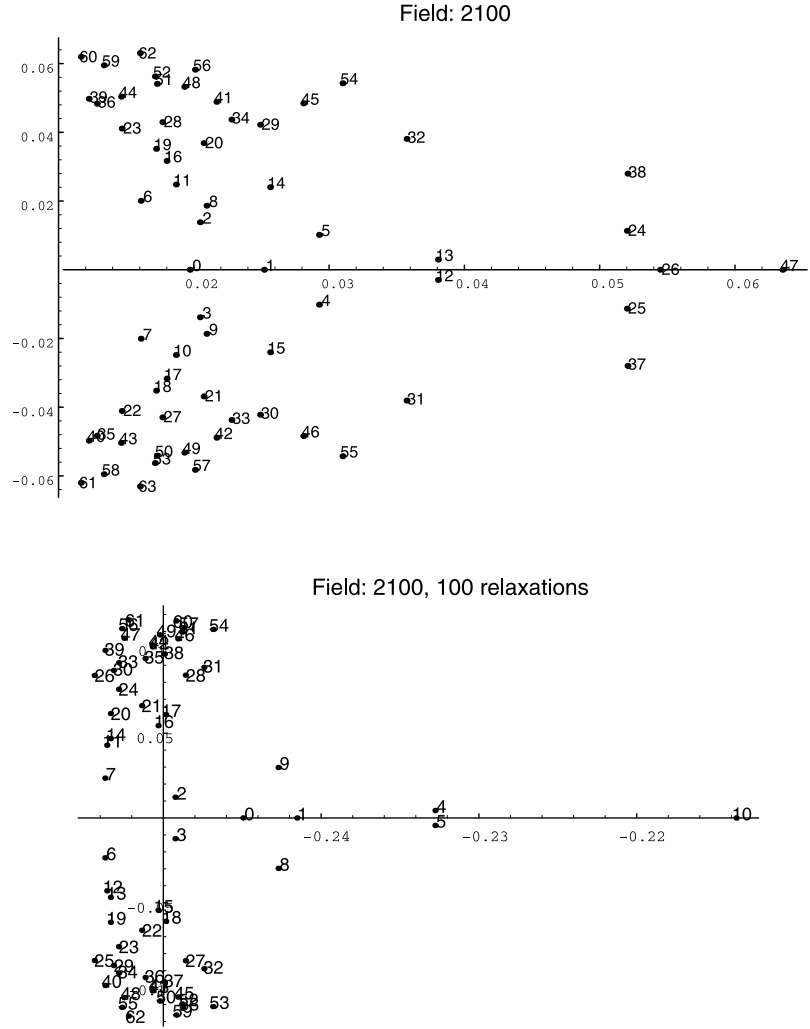 scaled 850}
\medskip
\caption{Lowest 64 complex eigenvalues of the Wilson--Dirac operator for an
unquenched gluon configuration both before (upper plot) and after cooling (lower
plot).  The scale is such that  0.06 on the imaginary
axis roughly corresponds to the lowest Matsubara frequency, 380 MeV.}
\label{F:JN:4}  
\end{center}
\end{figure}

\subsection{Eigenmodes of the Dirac operator}
In the continuum limit, the Dirac operator for Wilson fermions approaches the
familiar continuum result 
\begin{eqnarray}
D \psi_x &=& \psi_x - \kappa \sum_\mu \Bigl[ (r-\gamma_\mu) u_{x,\mu}
\psi_{x+\mu} + (r + \gamma_\mu) u^\dagger_{x-\mu,\mu} \psi_{x-\mu}
\Bigr]\nonumber\\
&&\quad
\to
\frac{1}{m} \left[ m+i (\notp + g \, \notA)\right] \psi\  .
\end{eqnarray}

In the free case, the continuum spectrum is $\frac{1}{m}[m+i |\vec{p}|]$ and the
Wilson lattice operator approximates this spectrum in the physical regime and
pushes the unphysical fermion modes to  very large (real) masses.   In the presence
of an instanton of size $\rho$ at $x=0$, it is shown in Ref.~\cite{R:JN:15} that the lattice
operator produces a mode with zero imaginary part that approaches the
continuum result 
\be
\psi_0(x)_{s,\alpha} = u_{s, \alpha} \frac{\sqrt{2}}{\pi} \frac{\rho}{(x^2 +
\rho^2)^{3/2}}
\ee
and whose mixing with other modes goes to zero as the lattice volume goes to
infinity.  In addition, instanton-anti-instanton pairs that interact sufficiently form
complex conjugate pairs of eigenvalues that move slightly off the real axis.  Thus,
by observing the Dirac spectrum for a lattice gluon configuration containing a
collection of instantons and anti-instantons, it is possible to identify zero modes
directly in the spectrum.

Fig.~\ref{F:JN:4} shows the lowest 64 complex eigenvalues of the Dirac operator on
a $16^4$ unquenched gluon configuration for ${6}/{g^2}=5.5$ and $\kappa=0.16$,
both before and after cooling (where 100 relaxation steps with a parallel algorithm
are comparable to 25 cooling steps).  The lower, cooled, plot has just the structure
we expect with a number of isolated instantons with modes on the real axis and
pairs of interacting instantons slightly off the real axis.  However, even though the
uncooled case shown in the upper plot also contains fluctuations several orders of
magnitude larger than the instantons (as seen in Fig.~\ref{F:JN:2}), it shows the
same structure of isolated instantons and interacting pairs.  To set the scale, note
that if we had antiperiodic boundary conditions in time, the lowest Matsubara
mode ($ip = i\frac{\pi}{L}$) would occur at 0.06 on the imaginary axis, so all the
modes below this value are presumably the results of zero modes.

\subsection{Zero mode expansion}
The Wilson--Dirac operator has the property that $D=\gamma_5 D^\dagger
\gamma_5$, which implies that $\langle \psi_j |\gamma_5| \psi_i \rangle=0$ unless
$\lambda_i = \lambda^*_j$ and we may write the spectral representation of the
propagator
$$
\langle x |D^{-1}|y \rangle = \sum_i \frac{\langle x|\psi_i \rangle \langle \psi_{\bar{i}}
|\gamma_5| y\rangle}{\langle \psi_{\bar{i}}|\gamma_5| \psi_i\rangle \lambda_i}
$$
where $\lambda_i = \lambda_{\bar{i}}^*$.  A clear indication of the role of zero modes
in light hadron observables is the degree to which truncation of the expansion to the
zero mode zone reproduces the result with the complete propagator.

\begin{figure}[h!] 
\begin{center}
\BoxedEPSF{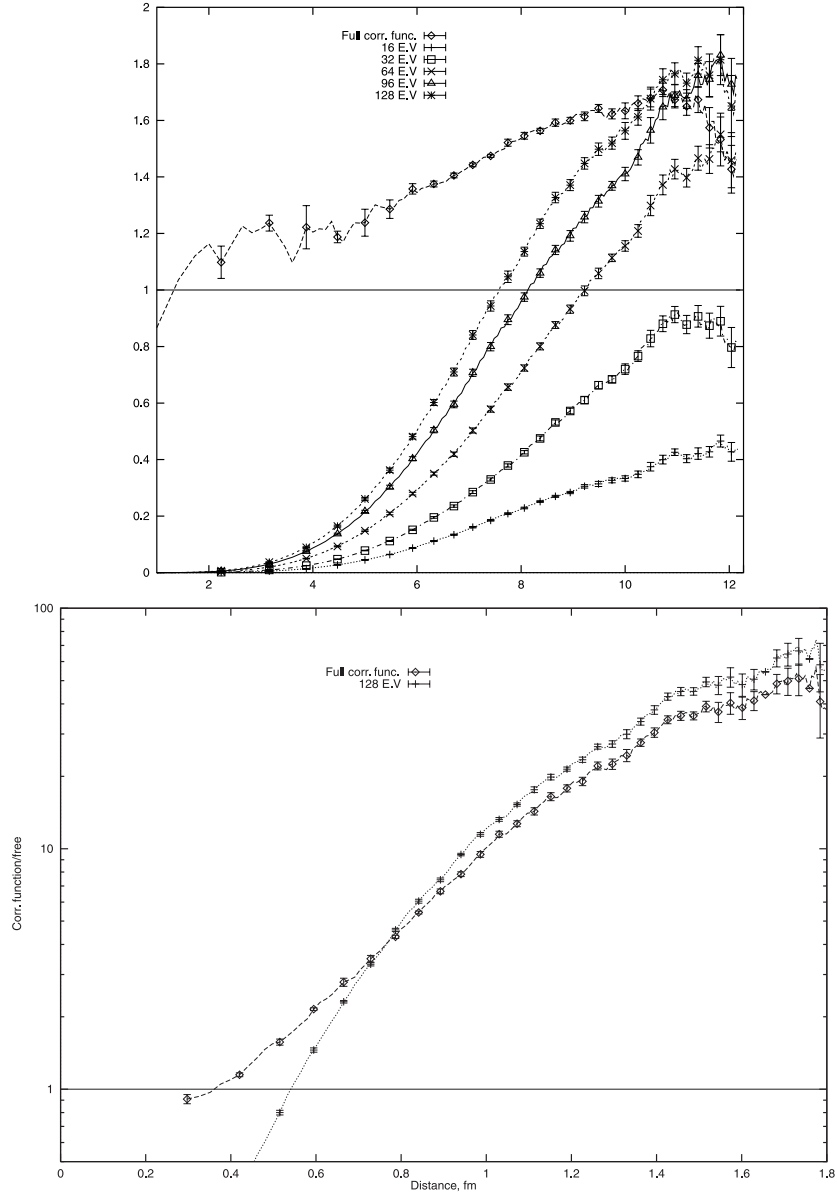 scaled 900}
\caption{Contributions of low Dirac eigenmodes to the vector (upper graph)
and pseudoscalar (lower graph) vacuum correlation functions.  The upper
graph shows the contributions of 16, 32, 64, 96, and 128 eigenmodes compared with
the full correlation function for an unquenched configuration with a 63~MeV valence
quark mass.  The lower graph compares 128 eigenmodes with the full correlation
function for a quenched configuration with a 23~MeV quark mass.}
\label{F:JN:5}  
\end{center}
\end{figure}

Fig.~\ref{F:JN:5} shows the result of truncating the vacuum correlation functions
for the vector and pseudoscalar channels to include only low
eigenmodes.\cite{R:JN:15} On a $16^4$ lattice, the full propagator contains
786,432 modes. The top plot of Fig.~\ref{F:JN:5}  shows the result of including the
lowest 16, 32, 64, 96, and finally 128 modes. Note that the first 64 modes
reproduce most of the strength in the rho resonance peak pointed out in
Fig.~\ref{F:JN:1}, and by the time we include the first 128 modes, all the strength
is accounted for.  Similarly, the lower plot in Fig.~\ref{F:JN:5} shows that the
lowest 128 modes also account for  the analogous pion contribution to the
pseudoscalar vacuum correlation function.  Thus, without having to resort to
cooling, by looking directly at the contribution of the lowest eigenfunctions, we have
shown that the zero modes associated with instantons dominate the propagation
of rho and pi mesons in the QCD vacuum.

\subsection{Localization}
Finally, it is interesting to ask whether the lattice zero mode eigenfunctions are
localized on instantons.  
  This was studied by plotting the quark density distribution for
  individual eigenmodes in
  the $x$-$z$ plane for all values of $y$ and~$t$, and comparing with
  analogous plots of the action density. As expected, for a cooled
  configuration the eigenmodes 
correspond to
linear combinations of localized zero modes at each of the instantons.
(Because there are no symmetries, the coefficients are much more
complicated than the even and odd combinations in a double well or the
Bloch waves in a periodic potential.)  What is truly remarkable, however, is
 that the eigenfunctions of the uncooled configurations also
exhibit localized peaks at locations at which instantons are identified by
cooling.  Thus, in spite of the fluctuations several orders of magnitude larger
than the instanton fields themselves, the light quarks essentially average
out these fluctuations and produce localized peaks at the topological
excitations.  

\section{Conclusion}\label{JWN:sec:6}
Taken as a whole, lattice calculations now provide strong evidence
that instantons play a dominant role in quark propagation in the vacuum
and in light hadron structure.  The instanton content of
gluon configurations has been extracted by cooling, and  the instanton size
and density are consistent with the instanton liquid model.  Vacuum correlation
functions, ground state density-density correlation functions, and masses calculated with
 only instantons show striking agreement with the results obtained with all gluons.
 Zero modes associated with instantons are clearly evident in the
Dirac spectrum, and  account for the rho and pi contributions to vector and pseudoscalar
vacuum correlation functions.  Finally,  quark localization at
instantons has been observed directly in uncooled configurations.  Hence, in
addition to providing a powerful tool for calculating a variety of important physical
observables, ranging from spectroscopy and weak matrix elements to QCD
thermodynamics, Lattice~QCD is also beginning to teach us the underlying physics of
hadron structure, bit by bit. 

\acknowledgments
It is a pleasure to thank Professor Alfredo Molinari and Professor Renato Ricci for
their hospitality at Varenna and
 to acknowledge the essential role of Richard Brower, Ming Chu, Jeff
Grandy, Suzhou Huang, Taras Ivanenko, Kostas Orginos, and Andrew Pochinsky who
collaborated in various aspects of this work.  We are also grateful for the donation by
Sun Microsystems of the 24 Gflops E5000 SMP cluster on which the most recent
calculations were performed and the computer resources provided by NERSC with
which this work was begun. This work is
supported in part by funds provided by the US Department of Energy (DOE) under
cooperative research agreement \#DF-FC02-94ER40818.


\begin{thebibliography}{00}
\def\J#1#2#3#4{{\it#1} {\bf#2}, #3 (#4)}


\bibitem{R:JN:03} Shuryak  E.V., \J{Rev. Mod. Phys.}{65}{1}{1993}, \J{Nucl.
Phys.}{B {\rm (Proc. Suppl.)} 34}{107}{1994}, and Sch\"affer T., and Shuryak E.V.,
hep-ph/9610451v2.

\bibitem{R:JN:04} Shuryak  E.V., and Verbaarschot J.J.M., \J{Nucl.
Phys.}{B410}{55}{1993}; Sch\"affer T., Shuryak E.V., and Verbaarschot J.J.M., \J{Nucl.
Phys.}{B412}{143}{1994}.
\bibitem{R:JN:05} Dyakanov  D.I., and Petrov  V.  Yu,  \J{Nucl. Phys.}{B245}{259}{1984};
\J{}{B272}{457}{1986}.

\bibitem{R:JN:01} Gross  D., and Wilczek  F., \J{Phys. Rev.}{D9}{980}{1974}.

\bibitem{R:JN:02} Ji X., Tang J., and Hoodbhoy P., \J{Phys. Rev. Lett.}{76}{740}{1996}.


\bibitem{M:JN:5a} Dashen R. and Gross D.J., \J{Phys. Rev.}{D23}{2340}{1981}.


\bibitem{M:JN:1} Hasenfratz P. and Niedermayer F., \J{Nucl.
Phys.}{B414}{785}{1994}; Bietenholz W. and Wiese U.-J., \J{Nucl.
Phys.}{B464}{319}{1996}.
\bibitem{R:JN:10} Negele  J.W., and Orland  H., {\it Quantum Many-Particle Systems},
New York: Addison-Wesley, 1987.

\bibitem{R:JN:06}  Creutz M, {\it Quarks, Gluons and Lattices},
Cambridge: Cambridge Univ Pr, 1983.

\bibitem{M:JN:xx}
{\it Lattice Gauge Theories and Monte Carlo Simulations},  Rebbi C., ed. Singapore:
World Scientific,  1983.

\bibitem{M:JN:2} 
Rothe H., {\it Lattice Gauge Theories: An Introduction}, World Scientific, 1992.

\bibitem{M:JN:3} 
Montvay I  and M\"unster  G, {\it Quantum Fields on a Lattice}, Cambridge: Cambridge
Univ Pr, 1994.

\bibitem{M:JN:12}
Sharpe S., in {\it Phenomenology and Lattice QCD}, Kilcup G. and Sharpe S., eds.
World Scientific, 1993.


\bibitem{M:JN:13}
Karsten L.H. and Smit J., \J{Nucl. Phys.}{B183}{103}{1981}.

\bibitem{M:JN:14}
Nielsen H.B. and Minomiya M., \J{Nucl. Phys.}{B185}{20}{1981}.

\bibitem{M:JN:15}
Kogut J. and Susskind L., \J{Phys. Rev.}{D11}{395}{1975}.

\bibitem{M:JN:16}
Kaplan D.B., \J{Phys. Lett.}{B288}{342}{1992};
Furman V. and Shamir Y.,  \J{Nucl. Phys.}{439}{54}{1995}.

\bibitem{R:JN:07} Chu  M.-C., Grandy  J.M., Huang  S., and Negele  J.W.,
\J{Phys.~Rev.~Lett.}{70}{225}{1993}; \J{Phys. Rev.}{D48}{3340}{1993}.

\bibitem{R:JN:08} \leavevmode\llap{'}t Hooft G.  \J{Phys. Rev.}{14D}{3432}{1976}.

\bibitem{R:JN:09} Belavin  A.A., Polyakov  A.M., Schwartz  A.P., and Tyupkin Y.S.
\J{Phys. Lett.}{59B}{85}{1975}.


\bibitem{R:JN:11}    Berg  B. \J{Phys. Lett.}{104B}{475}{1981}; Teper  M., \J{Nucl.
Phys.}{B {\rm (Proc. Suppl.)} 20}{159}{1991}.

\bibitem{R:JN:12} Chu  M.-C., Grandy  J.M., Huang  S., and Negele  J.W.,
\J{Phys. Rev.}{D49}{6039}{1994}.

\bibitem{R:JN:13} Chu  M.-C., Lissia  M., and Negele  J.W., \J{Nucl.
Phys.}{B360}{31}{1991}; Lissia  M., Chu  M.-C.,  Negele  J.W., and Grandy J.M., \J{Nucl.
Phys.}{A555}{272}{1993}.

\bibitem{R:JN:14} de Forcrand P., Garcia P\'erez M., and Stamatescu I.-O.,
hep-lat/9701012 (1997). 

\bibitem{R:JN:15} Ivanenko T., and Negele
J.W., {\it Proceedings of Lattice '97}, to be published in {\it Nucl. Phys.},
hep-lat/9709130;
 Ivanenko  T., MIT Ph.D. dissertation 1997.


\end{thebibliography}
\end{document}